\pdfoutput=1
\documentclass[preprint]{elsarticle-yjc}
\usepackage{lineno,hyperref}
\modulolinenumbers[5]
\newcommand{\remove}[1]{}

\usepackage{myMac-sgp12-yjc} 

\renewcommand{\draftdim}{
		\setlength{\oddsidemargin}{0.25in}   
		\setlength{\evensidemargin}{0.25in}  
                \setlength{\textwidth}{6in}
		}%
\draftdim

\newcommand{\sect}[2][]{\section{#2}\label{sec:#1}}
\newcommand{\ssect}[2][]{\subsection{#2}\label{ssec:#1}}

\newcommand{\myHrefxx}[2][blue]{
	    \href{#2}{\textcolor{#1}{\url{#2}}}}

 	\newcommand{\bproDIY}[3][0.15in]{
 		\vspace*{#1}
 		\noindent \textbf{Proposition #2.}
 		\textit{ #3 }\\
 	}

\newcommand{\Sep}{\ensuremath{\mathrm {Sep}}}	

\journal{Computational Geometry: Theory and Applications}

\newcommand{\blindOrNot}[2]{#2}             

	\newcommand{\wtC}{\wt{C}}		

	  \newcommand{\wtphi}{\wt{\phi}}	

	\newcommand{\free}{\ensuremath{\mathtt{FREE}}}
	\newcommand{\stuck}{\ensuremath{\mathtt{STUCK}}}
	\newcommand{\mixed}{\ensuremath{\mathtt{MIXED}}}

	\newcommand{\nopath}{{\tt NO-PATH}}

\remove{
} 

\begin{document}

\begin{frontmatter}

\title{Soft Subdivision Motion Planning for Complex Planar Robots\tnoteref{title-note-label}}
\tnotetext[title-note-label]{
 The conference version of this
 paper~\cite{zhou-chiang-yap:complex-robot:18} appeared in {\it
   Proc.\ 26th European Symposium on Algorithms (ESA 2018)}, pages
 73:1-73:14, 2018.  Helsinki, Finland, Aug.~20-24, 2018. 
This work is supported in part by NSF Grants \#CCF-1423228 and \#CCF-1563942.
}

\author[Tandon-address]{Bo Zhou}
        \ead{bz387@nyu.edu}
\author[Tandon-address]{Yi-Jen Chiang\corref{cor1}}
        \ead{chiang@nyu.edu}
        \cortext[cor1]{Corresponding author.}
\author[Courant-address]{Chee Yap}
        \ead{yap@cs.nyu.edu}
\address[Tandon-address]
        {Department of Computer Science and Engineering, New York University, 
         Brooklyn, NY, USA.}
\address[Courant-address]
        {Department of Computer Science, New York University, 
         New York, NY, USA.}

\remove{ 
\author{Bo Zhou}{Department of Computer Science and Engineering, New York University, 
                 Brooklyn, NY, USA; \texttt{bz387@nyu.edu}.}{}{}{} 
\author{Yi-Jen Chiang}{Department of Computer Science and Engineering, New York University, 
                        Brooklyn, NY, USA; \texttt{chiang@nyu.edu}.}{}{}{}
\author{Chee Yap}{Department of Computer Science, New York University, 
                        New York, NY, USA; \texttt{yap@cs.nyu.edu}.}{}{}{}
\authorrunning{B.\ Zhou, Y.-J.\ Chiang and C.\ Yap}
%

\Copyright{Bo Zhou, Yi-Jen Chiang and Chee Yap}

\subjclass{Theory of computation $\rightarrow$ Randomness,
  geometry and discrete structures $\rightarrow$ Computational
  geometry; Computing methodologies $\rightarrow$ Artificial
  intelligence $\rightarrow$ Planning and scheduling $\rightarrow$ Robotic
  planning.}
%


} 

%
\begin{abstract}
        The design and implementation of theoretically-sound robot
        motion planning algorithms is challenging.  Within the
        framework of {\em resolution-exact algorithms}, it is possible
        to exploit {\em soft predicates} for collision detection.  The
        design of soft predicates is a balancing act between easily
        implementable predicates and their accuracy/effectivity.

	In this paper, we focus on the class of planar polygonal rigid robots
	with arbitrarily complex geometry.  We exploit the remarkable
	{\em decomposability} property of soft collision-detection
	predicates of such robots.  We introduce a general
	technique to produce such a decomposition.
	If the robot is an $m$-gon, the complexity of this
	approach scales linearly in $m$.  This contrasts
	with the $O(m^3)$ complexity known for exact planners.
	It follows that we can now routinely produce soft predicates
	for any rigid polygonal robot.  This results in
	resolution-exact planners for such robots within the
	general {\em Soft Subdivision Search} (SSS) framework.
	This is a significant advancement in the theory of
	sound and complete planners for planar robots.

	We implemented such decomposed predicates in our open-source
        \blindOrNot{library}{\corelib}.  The experiments show that our
        algorithms are effective, perform in real time on non-trivial
        environments, and can outperform many sampling-based methods.

\end{abstract}


\end{frontmatter}

\sect[intro]{Introduction} \label{se-intro}
	Motion planning is widely studied in robotics
\cite{latombe:robot-motion:bk,lavalle:planning:bk,choset-etal:bk}.
	Many planners are heuristic, i.e., without a priori guarantees
	of their performance (see below for what we mean by guarantees).
	In this paper, we are interested in non-heuristic
	algorithms for the \dt{basic planning problem}:
	this basic problem considers only kinematics and the existence of paths.
	The robot $R_0$ is fixed,
	and the input is a triple $(\alpha,\beta,\Omega)$ where
	$\alpha, \beta$ are the start 
	and goal configurations of $R_0$, 
	and $\Omega\ib\RR^d$ is a polyhedral environment in $d=2$ or $3$.
	The algorithm outputs
	an $\Omega$-avoiding path from $\alpha$ to $\beta$ if one
	exists, and  \nopath\ otherwise.  See \refFig{robot} for
	some 
	rigid robots, and also \refFig{gui} for our GUI interface
	for path planning.

%
\begin{figure}[t]
 \begin{center}
   \begin{minipage}[t]{0.45\linewidth}
      \includegraphics[scale=0.12]{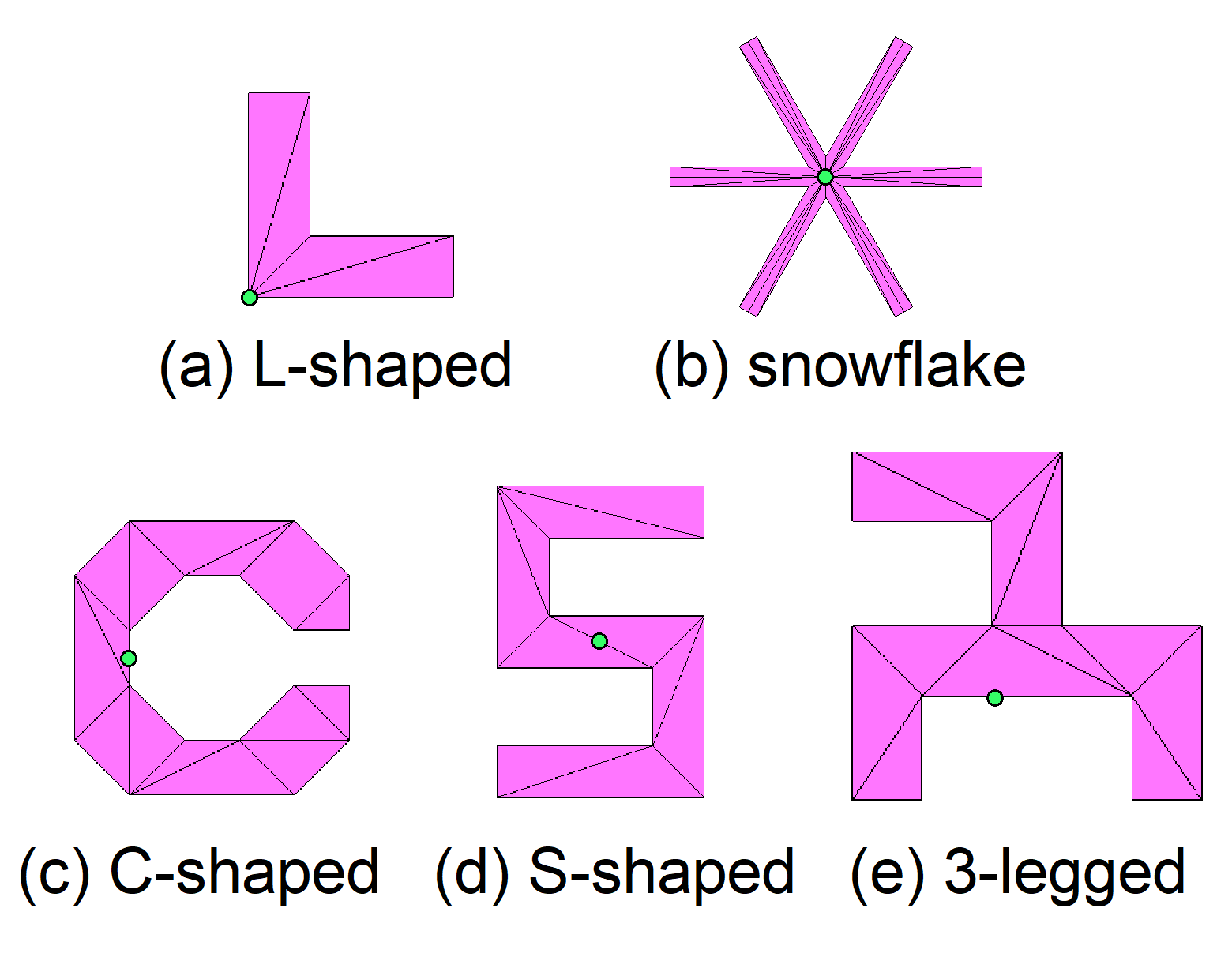} 
      \caption{Some rigid planar robots ((a)-(b): star-shaped; (c)-(e): general shaped). 
               \label{fig:robot}}
   \end{minipage} \quad\quad
   \begin{minipage}[t]{0.45\linewidth}
      \includegraphics[scale=0.14]{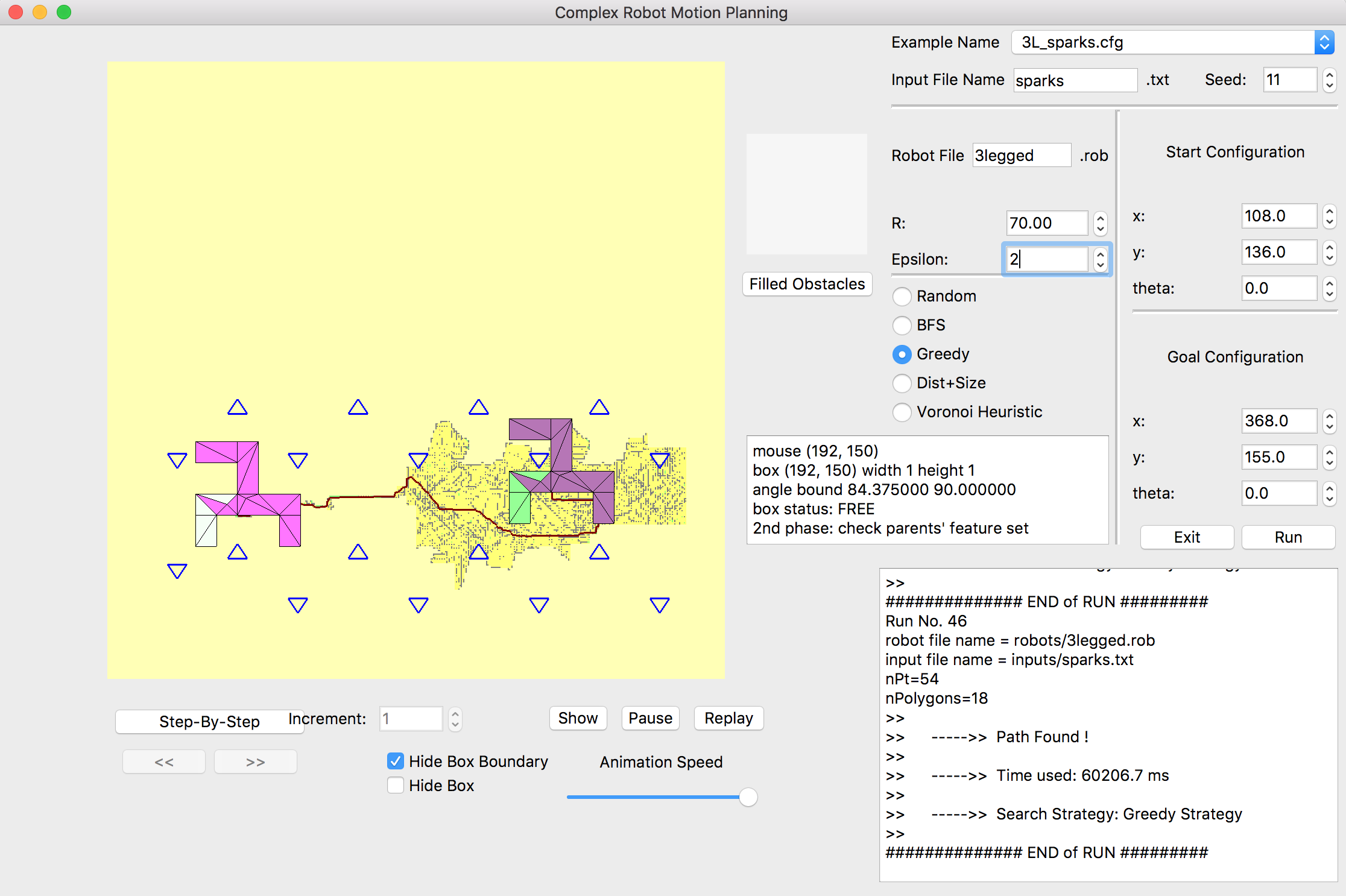}
      \caption{GUI interface for planner for a 3-legged robot. \label{fig:gui}}
   \end{minipage}
 \end{center}
\end{figure}

\remove{ 
\begin{figure}[b]
\begin{center}
\includegraphics[scale=0.15]{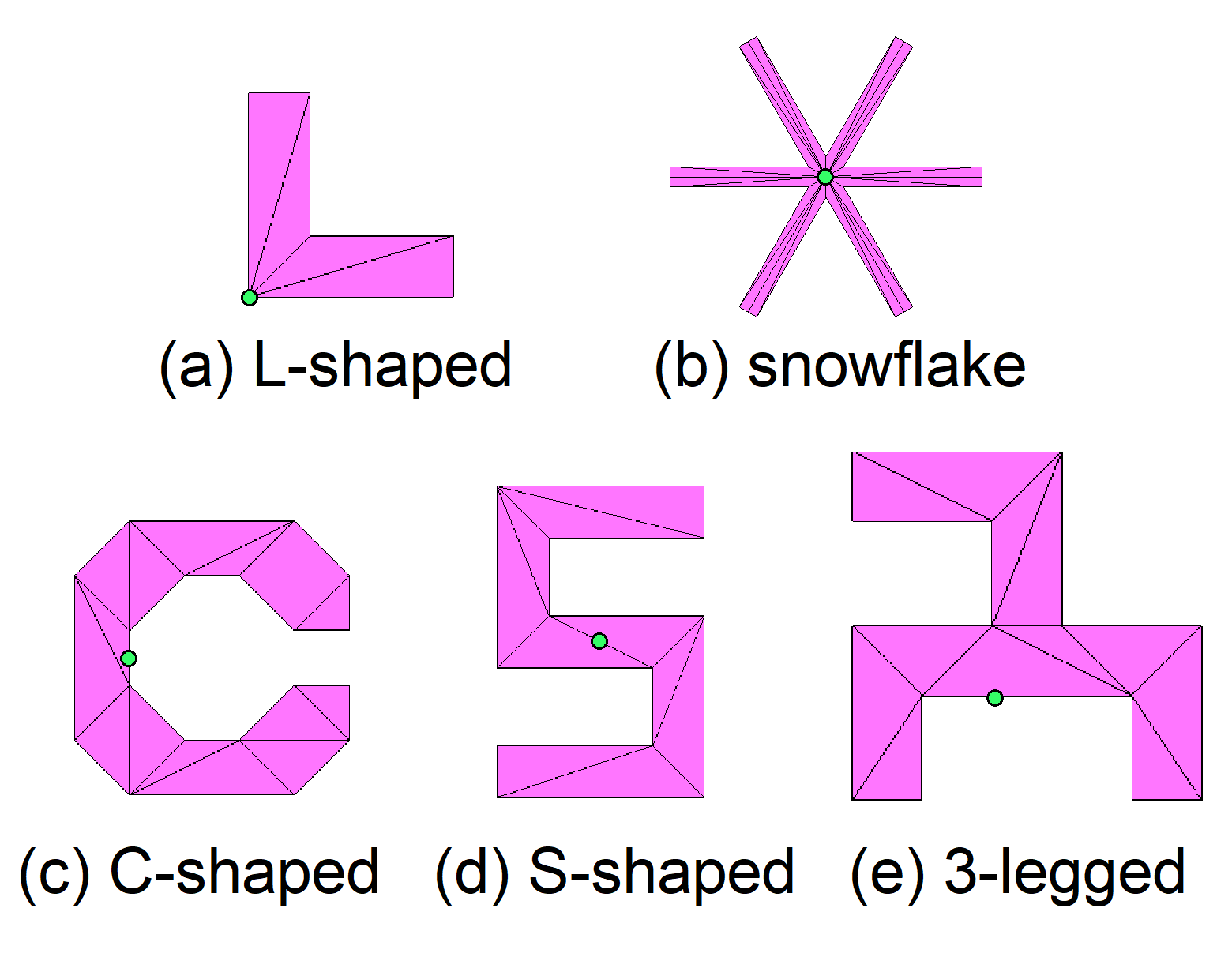} 
\caption{Some rigid planar robots. \label{fig:robot}}
\end{center}
\end{figure}

\begin{figure}[b]
\begin{center}
\includegraphics[scale=0.2]{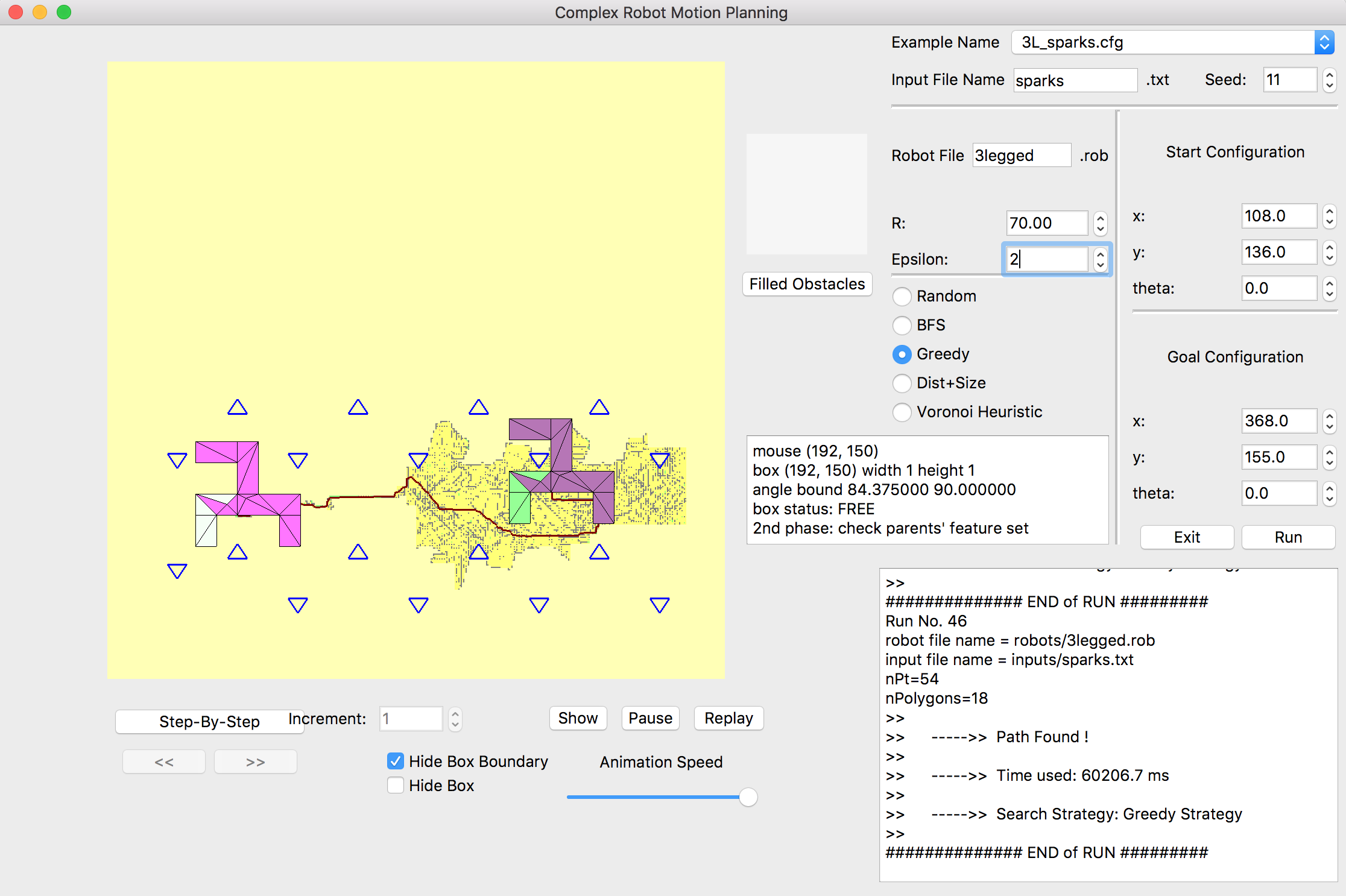}
\caption{GUI interface for planner for a 3-legged robot. \label{fig:gui}}
\end{center}
\end{figure}
} 

\remove{
 	\begin{figure*}
		    \begin{center}
                       \begin{minipage}[b]{0.45\linewidth}
			\includegraphics[scale=0.4]{figs/complex-robot2}
			\caption{Some rigid planar robots \label{fig:robot}}
		      \end{minipage} \quad\quad
	  	    \begin{minipage}[t]{0.45\linewidth}
			\includegraphics[scale=0.14]{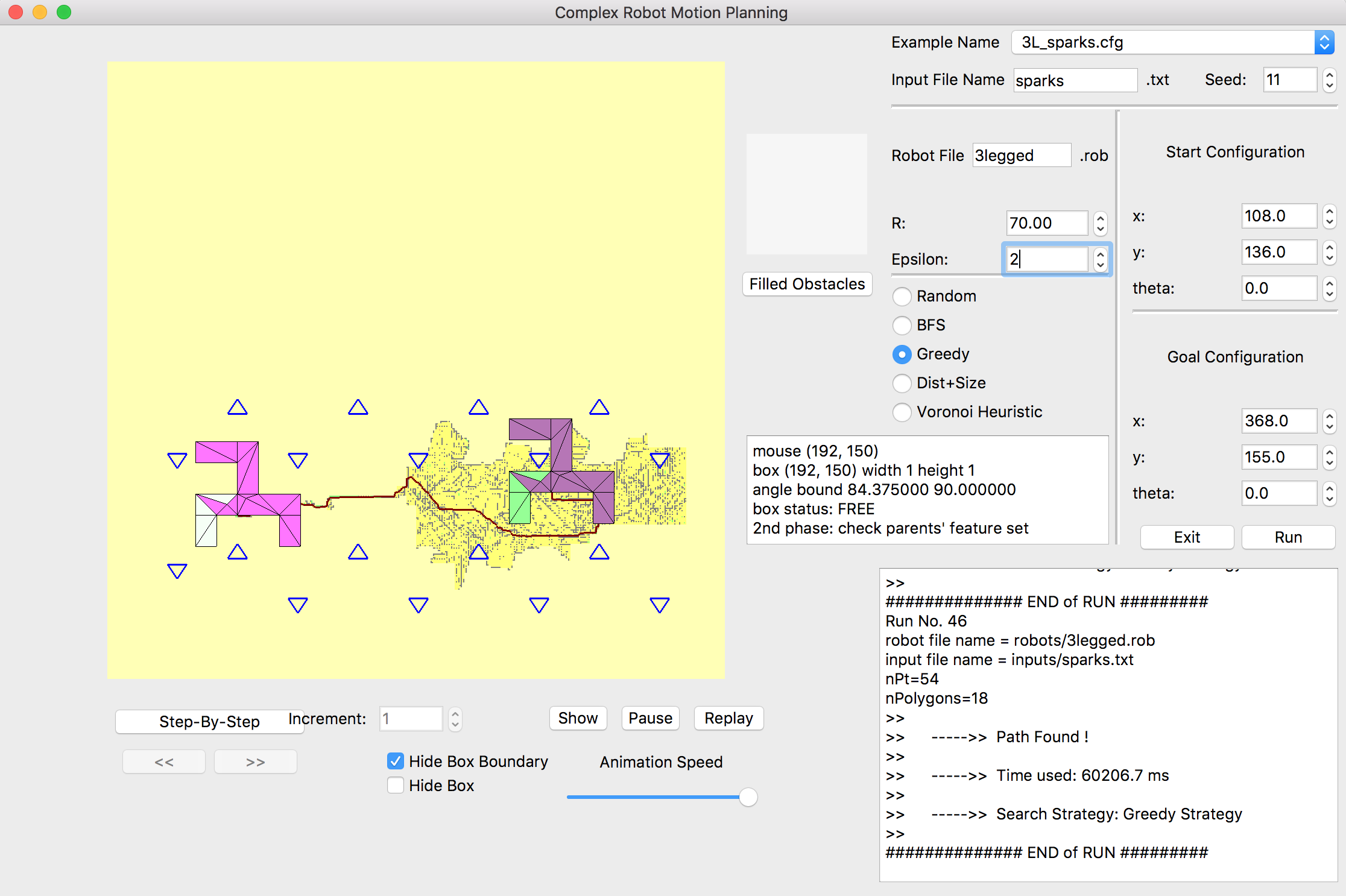}
			\caption{GUI interface for planner for
				snowflake robot\label{fig:gui}}
		    \end{minipage}
	  	  \end{center}
  	\end{figure*}
} 

	The basic planning problem
	ignores issues such as the optimality of paths, robot
	dynamics, planning in the time dimension,
	non-holonomic constraints, and other considerations
	of a real scenario.  Despite such an idealization,
	the solution to this basic planning problem is often
	useful as the basis for finding solutions that do take into account
	the omitted considerations.  E.g., given a kinematic path,
	we can plan a smooth trajectory with a homotopic trace.

	The algorithms for this basic problem are called ``planners.''
	In theory, it is possible to design {\em exact} planners
	because the basic path planning
	is a semi-algebraic (non-transcendental) problem.
	Even when such algorithms are available, exact planners
	have relatively high complexity and are non-adaptive,
	even in the plane (see
	\cite{milen-sacks-trac:plane-planner:13}).
	So we tend to see inexact implementations of exact algorithms,
	with unclear guarantees.
	%
	%
	\ignore{ Cristian Calude, editor. 
	  The Human Face of Computing, chapter Kurt Mehlhorn: From Theory to
	  Library of Efficient Data Types and Algorithms (LEDA) and Algorithm
	  Engineering, pages 59-72.  Imperical College Press, 2015
	}%
	When fully explicit algorithms are known,
	exact implementation of exact planners is possible using
	suitable software tools such as 
	the \cgal\ library \cite{halperin-fogel-wein:bk}.

	In current robotics \cite{lavalle:planning:bk,choset-etal:bk},
	those algorithms that are considered practical and
 	have some guarantees may be classified as either resolution-based
	or sampling-based.  
	The guarantees for the former is
	the notion of \dt{resolution completeness} and for the latter,
	\dt{sampling completeness}.  Roughly speaking, 
	{\em if there exists a path} then:
	\\ -- resolution completeness says that a path will be found if
	the resolution is fine enough;
	\\ -- sampling completeness says that a path will be found with high
	probability if ``enough'' random samples are taken.
	\\ But notice that if there is no path, these criteria
	are silent; indeed, such algorithms would not halt except
	by artificial cut-offs.
	Thus a major effort in the last 20 years of sampling
	research has been devoted to the so-called ``Narrow Passage''
	problem. It is possible to
	view this problem as a manifestation of the \dt{Halting Problem} for
	the sampling approaches: how can the
	algorithm halt when there is no path? 
	%
	(A possible approach to address this problem might be to combine
	sampling with exact computation, as in~\cite{Halperin-ESA11}.)
	
	Motivated by such issues, as well as trying to
	avoid the need for exact computation, 
\blindOrNot{the authors}{we} 
        in \cite{wang-chiang-yap:motion-planning:15,yap:sss:13}
	introduced the following replacement for resolution
	complete planners:
	a \dt{resolution-exact planner}
	takes an extra input parameter $\eps>0$
	in addition to $(\alpha,\beta,\Omega)$,
	and it always halts and outputs either an $\Omega$-avoiding
	path from $\alpha$ to $\beta$ or \nopath.
	The output satisfies this condition: there is a constant
	$K>1$ depending on the planner, but independent of the inputs,
	such that:
	\\ -- if there is a path of clearance $K\eps$, it must output
	a path;
	\\ -- if there is no path of clearance $\eps/K$, it
	must output \nopath.   
	\\ Notice that if the optimal clearance
	lies between $K\eps$ and $\eps/K$, then the algorithm
	may output either a path or \nopath. 
	So there is output indeterminacy.
	Note that the traditional way of using $\eps$
	is to fix $K=1$, killing off indeterminacy.
	Unfortunately, this also leads us right back to exact
	computation which we had wanted to avoid. 
	%
	%
	We believe that indeterminacy is a small price to pay in exchange for
	avoiding exact computation~\cite{wang-chiang-yap:motion-planning:15}.
	The practical efficiency of resolution-exact algorithms is
	demonstrated by implementations of planar robots with 2, 3
	and 4 degrees of freedom (DOF)
        \cite{wang-chiang-yap:motion-planning:15,
	luo-chiang-lien-yap:link:14,yap-luo-hsu:thicklink:16},
	and also 5-DOF spatial robots
	\cite{hsu-chiang-yap:rod-ring:19}.
	All these robots perform in real-time in non-trivial environments.
	In view of the much stronger guarantees of performance,
	resolution-exact algorithms might reasonably be expected
	to have a lower efficiency compared to sampling algorithms.
	Surprisingly, no such trade-offs were observed: resolution-exact
	algorithms consistently outperform
	sampling algorithms.  Our 2-link robot
	\cite{luo-chiang-lien-yap:link:14,yap-luo-hsu:thicklink:16}
	was further generalized to have thickness (a feat that exact methods
	cannot easily duplicate), and can satisfy a non-self-crossing
	constraint, all without any appreciable slowdown. 
	Finally, these planners are more general than
	the basic problem: they all work for parametrized families
	$R_0(t_1,t_2\ldots)$ of robots, where $t_i$'s are robot parameters.
	All these suggest the great promise of our approach.
	

	{\bf What is New in This Paper.}
	In theoretical path planning,
	the algorithms often 
  considered
%
%
        simple robots
	like discs or line segments.
	In this paper, we 
  consider robots of complex shape, which are more realistic models
  for real-world robots.  We call them ``complex robots'' (where
  the complexity comes from the robot geometry rather than from the
  degrees of freedom).
%
%
	We focus on planar robots that are rigid and connected.
	Such a robot can be represented by a compact connected polygonal set
	$R_0\ib \RR^2$ whose boundary is an $m$-sided polygon, i.e.,
	an $m$-gon. 
%
	Informally, we call $R_0$ a ``complex robot''
	if it is a non-convex $m$-gon for ``moderately large'' values of $m$,
	say $m\ge 5$.
	By this criterion, all the robots in \refFig{robot} are ``complex.''
	According to
	\cite{zhang-kim-manocha:path-non-existence:08},
	no exact algorithms for $m>3$ have been implemented;
	in this paper, we have robots with $m=18$.
	To see why complex robots may be challenging, recall that
	the free space of such robots may have complexity
	$O((mn)^3\log(mn))$ (see
	\cite{avnaim-boissonnat-faverjob:practical-polygonal:89})
	when the robot and environment have complexity $m$ and $n$,
	respectively.
	Even with $m$ fixed, this can render the algorithm
	impractical.  For instance, if $m=10$, the algorithm may
	slow down by 3 orders of magnitude.
	But our subdivision approach does not
	have to compute the entire free space before planning a path;
	hence the worst-case cubic complexity of the free space is not
	necessarily an issue. 
	
	More importantly, we show that the complexity of our new method grows
	only linearly with $m$.  To achieve this, we
	exploit a remarkable property of
	soft predicates called ``decomposability.''  We show how
	an arbitrary complex robot can be decomposed (via triangulation
	that may introduce new vertices) into an ensemble of
	``nice triangles'' for which soft predicates are easy to implement. 
	As we see below, there is a significant difference between
	a single triangle and an ensemble of triangles.
	\ignore{
	In
	\cite{wang-chiang-yap:motion-planning:15},
	they choose the circumcenter of the triangle as the origin
	of the coordinate system to simplify the predicates.
	But for us, all the triangles in the ensemble
	must share the same coordinate system.  This complicates
	our predicates.  
	}%
	{\em In consequence of our new techniques, we can now routinely
	construct resolution-exact planners for any reasonably complex robot
	provided by a user.}   This could
	lead to a flowering of experimentation algorithmics
	in this subfield.

%
Technically, it is important to note that the previous soft predicate
construction for a triangle robot in
\cite{wang-chiang-yap:motion-planning:15,sss2} requires that the
rotation center, i.e., the origin of the (rotational) coordinate
system, be chosen to be the circumcenter of the triangle. But for our new
soft predicates the triangles in the triangulation of the complex
robot cannot be treated in the same way.  This is because all the
triangles of the triangulation must share a 
{\bf common origin}, 
%
%
to serve as the rotation center of the robot.
To ensure easy-to-compute predicates, we
introduce the notion of a ``nice triangulation''
relative to a chosen origin: all triangles must be ``nice'' relative to this
origin.  These ideas apply for arbitrary complex robots, but we also
exploit the special case of star-shaped robots to achieve stronger results.
%

%
	\refFig{gui} shows our experimental setup for complex robots.
	A demo showing the real-time performance of our
	algorithms is found 
        in the video clip available through this  
        web link:
%
%
%
\myHrefxx{https://cs.nyu.edu/exact/gallery/complex/complex-robot-demo.mp4}.
%
%
%

{\bf Remark.}
	Although it is not our immediate concern to address noisy environments
	and uncertainties, it is clear that our work can
	be leveraged to address these issues.
	E.g., users can choose $\eps>0$ to be correlated
	with the uncertainty in the environment and the precision of
	the robot sensors.   By using weighted Voronoi diagrams
	\cite{bennett-papadopoulou-yap:minimization:16},
	we can achieve practical planners that have
	obstacle-dependent clearances
	(larger clearance for ``dangerous'' obstacles). 

	\ignore{
	this means that exact computation is possible \cite{sharma-yap:crc}.
	The direct representation of algebraic numbers is impractical, but
	implicit representation in such systems as
	\corelib\ \cite{core2} or in \cgal\
	\cite{halperin-fogel-wein:bk} may be used.  
	But of the planners that do have guarantees, 
	success.  

	Theoretically-sound planners have been designed as far back
	as the 1980s.  But in the rare cases when these exact
	planners are implemented, the implementation
	are not guaranteed to be correct because of numerical issues.
	The exceptions
	are those algorithms implemented using the principles of
	Exact Geometric Computation such as in CGAL
	\cite{halperin-fogel-wein:bk}.  We introduced the notion
	of resolution-exactness in
	\cite{wang-chiang-yap:motion-planning:15} to side-step the need for
	exact computation: purely numerical approximation
	such as provided in BigNumber packages suffices
	to produce correct implementations.

	In this paper, we address another gap between theory and
	implementations.  Exact theoretical algorithms for planar
	rigid robots has been restricted to ``simple'' robots, e.g.,
	$n=2$ (rod \cite{xxx}),
	$n=3$ (triangle \cite{xxx,boissonnat})
	or $n=12$ (cross \cite{xxx}).
	In practice, planners for complex robots have been designed
	and implemented (e.g., the gear robot \cite{manocha}).
	In this paper, we demonstrate techniques for achieving
	resolution-exact planners for complex robots such as
	\refFig{complex}.
	}


	{\bf Previous Related Work.}
	An early work is
	Zhu-Latombe \cite{zhu-latombe:hierarchical:91}
	who also classify boxes into \free\ or \mixed\ or \stuck\ 
	\blindOrNot{}{(using our terminology below)}.
	They introduced the concept of 
	\dt{M-channels} (comprised of \free\ or \mixed\ leaf boxes),
	as a heuristic basis to find an F-channel comprising
	only of \free\ boxes.   Subsequent researchers
	(Barbehenn-Hutchinson
	\cite{barbehenn-hutchinson:single-source:95}
	and
	Zhang-Manocha-Kim \cite{zhang-kim-manocha:path-non-existence:08})
	continued this approach.  
	Researchers in resolution-based approaches were interested in
	detecting the non-existence of paths, but
	their solutions remain partial because they do not guarantee
	to always detect non-existence of paths (of sufficient clearances)
	\cite{basch+3:disconnection:01,zhang-kim-manocha:path-non-existence:08}.
	The challenge of complex robots was taken up by Manocha's
	group who implemented a series of such examples
        \cite{zhang-kim-manocha:path-non-existence:08}:
	a ``five-gear'' robot,
	a ``2-D puzzle'' robot 
	a certain ``star'' robot with 4 DOFs,
	and a ``serial link'' robot with 4 DOFs.
	Except for the ``star,'' the rest are planar robots.

	{\bf Overview of the Paper.}
        Section~\ref{se-review} reviews the fundamentals of our soft
        subdivision approach.  Sections~\ref{se-star-robot}
        and~\ref{se-general} describe our new techniques for
        star-shaped robots and for general complex robots,
        respectively. We present the experimental results in
        Section~\ref{se-results}, and conclude in
        Section~\ref{se-conclude}.
        All proofs are put in the appendix at the end of the paper.
        The conference version of this paper appeared
        in~\cite{zhou-chiang-yap:complex-robot:18}.


\sect[review]{Review: Fundamentals of Soft Subdivision Approach}
\label{se-review}
%
Our soft subdivision approach includes the following three fundamental
concepts
(see~\cite{wang-chiang-yap:motion-planning:15}
and the Appendix of~\cite{luo-chiang-lien-yap:link:14} for the
details):
\bitem
\item Resolution-exactness. 
This is an alternative replacement for the standard concept
of ``resolution completeness'' in the subdivision literature.
Briefly, a planner is
\dt{resolution-exact} if there is a constant $K>1$ such that if there is a
path of clearance $K \eps$, it will return a path, and if there is
no path of clearance $\eps/K$, it will return \nopath.  Here,
$\eps>0$ is an additional input to the planner, in
addition to the normal parameters.

\item Soft Predicates. 
Let $\intbox
\RR^d$ be the set of closed axes-aligned boxes in $\RR^d$.  
We are interested in predicates that classify boxes.
Let
$C:\RR^d\to\set{+1,0,-1}$ be an (exact) predicate where $+1,-1$ are
called definite values, and $0$ the indefinite value.
For motion planning, we may also
identify $+1/-1/0$ with \free/\stuck/\mixed, respectively.
In our application, if $p$ is a free configuration, then $C(p)=\free$;
if $p$ is on the boundary of the free space, $C(p)=\mixed$;
otherwise $C(p)=\stuck$.
We extend $C$ to boxes $B\in\intbox\RR^d$ as follows: for a definite
value $v\in\set{+1,-1}$, $C(B)\as v$ if $C(x)=v$ for every $x\in B$.
Otherwise, $C(B)\as 0$.  Call $\wtC:\intbox\RR^d\to \set{+1,0,-1}$ a
``soft version'' of $C$ if whenever $\wtC(B)$ is a definite value,
$\wtC(B)=C(B)$, and moreover, if for any sequence of boxes $B_i$
($i\ge 1$) that converges monotonically to a point $p$,
$\wtC(B_i)=C(p)$ for $i$ large enough.

\item Soft Subdivision Search (SSS) Framework.  
This is a general framework for a broad class of motion planning
%
%
algorithms.
%
%
One must supply a small number of subroutines with fairly general
properties in order to derive a specific algorithm.
%
%
For SSS, we need a predicate to classify boxes in the configuration space
as \free/\stuck/\mixed, a method to split boxes, a method to test
if two \free\ boxes are connected by a path of \free\ boxes, and a
method to pick \mixed\ boxes for splitting.  The power of such
frameworks is that we can explore a great variety of techniques and
strategies.  \blindOrNot{The power of such frameworks
	is seen in the benefits that similar frameworks
	have brought to the sampling research community.}{
    	Indeed we introduced the SSS framework
	to emulate such properties found in the sampling framework.}

\eitem

{\bf Feature-Based Approach.}
Following our previous work
~\cite{wang-chiang-yap:motion-planning:15,luo-chiang-lien-yap:link:14},
our computation and predicates are "feature based" whereby the
evaluations of box primitives are based on a set $\wtphi(B)$ of
features associated with the box $B$.
Given a polygonal set $\Omega\ib\RR^2$ of obstacles, the boundary
$\partial\Omega$ may be subdivided into a unique set of \dt{corners}
(points) and \dt{edges} (open line segments), called the \dt{features}
of $\Omega$.  Let $\Phi(\Omega)$ denote this feature set.  Our
representation of $f\in \Phi(\Omega)$ ensures this \dt{local property
  of $f$}: {\em for any point $q$, if $f$ is the closest feature to
  $q$, then we can decide if $q$ is inside $\Omega$ or not}.
To see this, first note that if $f$ is a corner, then $q$ is outside $\Omega$
iff $f$ is a convex corner of $\Omega$.
But if $f$ is an edge, our representation 
assigns an orientation to $f$ such that $q$ is inside $\Omega$ iff
$q$ lies to the left of the oriented line through $f$.

\sect[star]{Star-Shaped Robots} \label{se-star-robot}
%
We first consider star-shaped robots. A star-shaped region $R$ is one
for which there exists a point $A \in R$  such that any line through $A$
intersects $R$ in a single line segment.  We call $A$ a
\dt{center} of $R$.  Note that $A$ is not unique. 
When a robot $R_0$ is a star-shaped polygon, we
decompose $R_0$ into a set of triangles that share a {\bf common
vertex} at a center $A$. The rotations of the robot $R_0$ about the point
$A$ can then be reduced to the rotations 
of ``nice'' triangles about $A$.  The soft predicates of
nice triangles will be easy to implement because their footprints
have special representations.

\subsection[nice]{Nice Shapes for Rotation} 

	    \begin{figure}[htb]
	    	  \begin{center}
		   \scalebox{0.29}{
	    	     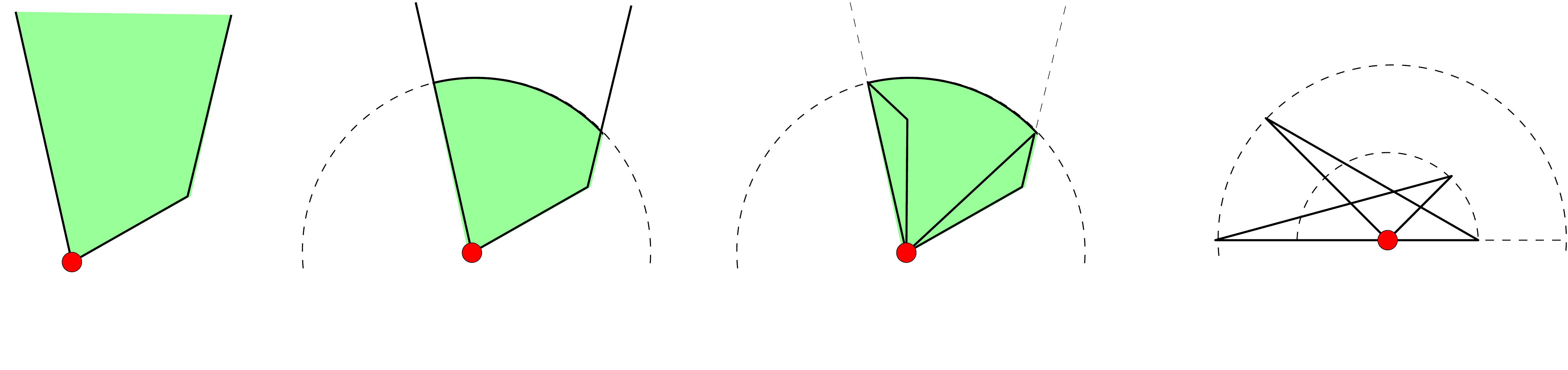}
	    	   \caption{Truncated triangular set and swept areas.}
	    	   \label{fig:generalized-tri-b}
	    	  \end{center}
	    \end{figure} 	 

        From now on, by a \dt{triangular set} we mean a subset
        $T\ib\RR^2$ which is written as the non-redundant intersection
        of three closed half-spaces: $T=H_1\cap H_2\cap H_3$.
        Non-redundant means that we cannot express $T$ as the
        intersection of only two half-spaces.
	Note that if $T$ is bounded, this is our familiar notion of
	a triangle with 3 vertices. But $T$ might be unbounded
	and have only 2 vertices as in \refFig{generalized-tri-b}(a).
	If $T$ is a triangular set, we may arbitrarily call one of its
	vertices the \dt{apex} and call the resulting $T$ a
	\dt{pointed triangular set}.
	By a \dt{truncated triangular set (TTS)}, we mean the intersection of
	a pointed triangular set $T$ with any disc centered at its apex $A$,
	as shown in \refFig{generalized-tri-b}(b). 

        
	\dt{Notation for Angular Range:}
	It is usual to identify $S^1$ (unit circle) with
	the interval $[0,2\pi]$ where $0$ and $2\pi$ are identified.
	Let $\alpha\neq \beta\in S^1$.  Then $[\alpha,\beta]$
	denote the range of angles from $\alpha$
	counter-clockwise to $\beta$.  Thus $[\alpha,\beta]$
	and $[\beta,\alpha]$ are complementary ranges in $S^1$.
	If $\Theta=[\alpha,\beta]$, then its \dt{width},
	$|\Theta|$ is defined as $\beta-\alpha$ if $\beta>\alpha$, and
	$2\pi+\beta-\alpha$ otherwise.
	Moreover, we will write
	``$\alpha<\theta<\beta$'' to mean that $\theta\in[\alpha,\beta]$.

	Fix an arbitrary bounded triangular set $T_0$, represented
	by its three vertices $A,B,C$ where $A$ is the apex.
	For $\theta\in S^1$, let $T_0[\theta]$ denote the
	footprint of $T_0$ after rotating $T_0$ counter-clockwise (CCW)
	by $\theta$ about the apex. 
	If $\Theta\ib S^1$, we write 
		$T_0[\Theta]=\bigcup\set{T_0[\theta]: \theta\in\Theta}$.
	The sets $T_0[\theta]$ and $T_0[\Theta]$ are called
	\dt{footprints} of $T_0$ at $\theta$ and $\Theta$, respectively.
	If $\Theta=[\alpha,\beta]$, write $T_0[\alpha,\beta]$
	for $T_0[\Theta]$, and call
	$T_0[\alpha,\beta]$ the \dt{swept area} as $T_0$ rotates
	from $\alpha$ to $\beta$.

	One of our concerns is to ensure that the
	swept area $T_0[\Theta]$ is ``nice.''  Consider
	an example where $[A,B,C]$ is a triangular set
	with apex $A$ (see \refFig{generalized-tri-b}(c)). 
	Consider the area swept by rotating $[A,B,C]$ in a CCW direction
	about its apex to position $[A,B',C']$.
	This sweeps out the truncated triangular set shown
	in \refFig{generalized-tri-b}(b).
	This truncated triangular set (TTS) is desirable
	since it can be easily specified by the intersection of
	three half-spaces and a disc.
	On the other hand, if $[A,B,C]$ is the triangular set in
	\refFig{generalized-tri-b}(d), then no rotation of $[A,B,C]$
	would sweep out a truncated triangular set.
	So the triangular set in 
	\refFig{generalized-tri-b}(d) is ``not nice,''
	unlike the triangular set in \refFig{generalized-tri-b}(c).

	In general, let $T=[A,B,C]$ be a bounded triangular set. 
	Let $a,b,c$ denote the corresponding angles at $A,B,C$.
	We say $T$ is \dt{nice}
	if either $b$ or $c$ is at least $\pi/2$ ($=90\degree$).
	We call the corresponding vertex ($B$ or $C$) a
	\dt{nice vertex}.  Assuming $T$ is non-degenerate
	and nice, there is a unique nice vertex.
	In the following, we assume (w.l.o.g.) that $B$ is the nice vertex.
	The reason for defining niceness is the following.

  	\blem \label{lem:nice-tri-set}
{\em
		Let $T$ be a pointed triangular set.
		Then $T$ is nice iff
		for all $\alpha\in S^1$ ($0<\alpha<\pi-a$),
		the footprints
		$T[0,\alpha]$ and $T[-\alpha,0]$
		are truncated triangular sets (TTS).
}
	\elem

\remove{
	\bpf
	If $T$ is nice, $T[0,\alpha]$ and $T[-\alpha,0]$ are clearly
        truncated triangular sets (TTS).  
        Conversely, if $T$ is not nice,
        let us assume that $\|A-B\|\le \|A-C\|$ (e.g.,
        \refFig{generalized-tri-b}(d)).  We claim that for
        sufficiently small $\alpha>0$, either $T[0,\alpha]$ or
        $T[-\alpha,0]$ is not a TTS.
	Assume (w.l.o.g.) that $A,B,C$ are in CCW order;
	we show that $T[0,\alpha]$ is not a TTS.
%
%
%

	If $T$ is not nice, then $b<90\degree$. Let $B-C$ intersects the
	$Circle B$ (the circle centered at $A$ that passes through $B$) 
	at $D$. Let $\alpha_{max} = \angle BAD
	= 180\degree-2b = 2(90\degree-b)$, 
since $b = \angle ABD = \angle ADB$.
        Note that a TTS is a {\bf convex} set as it is the intersection
        of three half-spaces and one disc; all of them are convex and thus
        the intersection is also convex.
	However, for any $\alpha < \alpha_{max}$, $T[0,\alpha]$ is not a TTS
	since $B-C$ will intersect $B'-C'$ inside $Circle B$ (see
	\refFig{not-nice-tri-set} for an example) --- 
  this makes $T[0,\alpha]$ {\bf non-convex} and thus it is not a TTS.
%
%
%
	\epf


	    \begin{figure}[htb]
	    	  \begin{center}
		   \scalebox{0.3}{
	    	     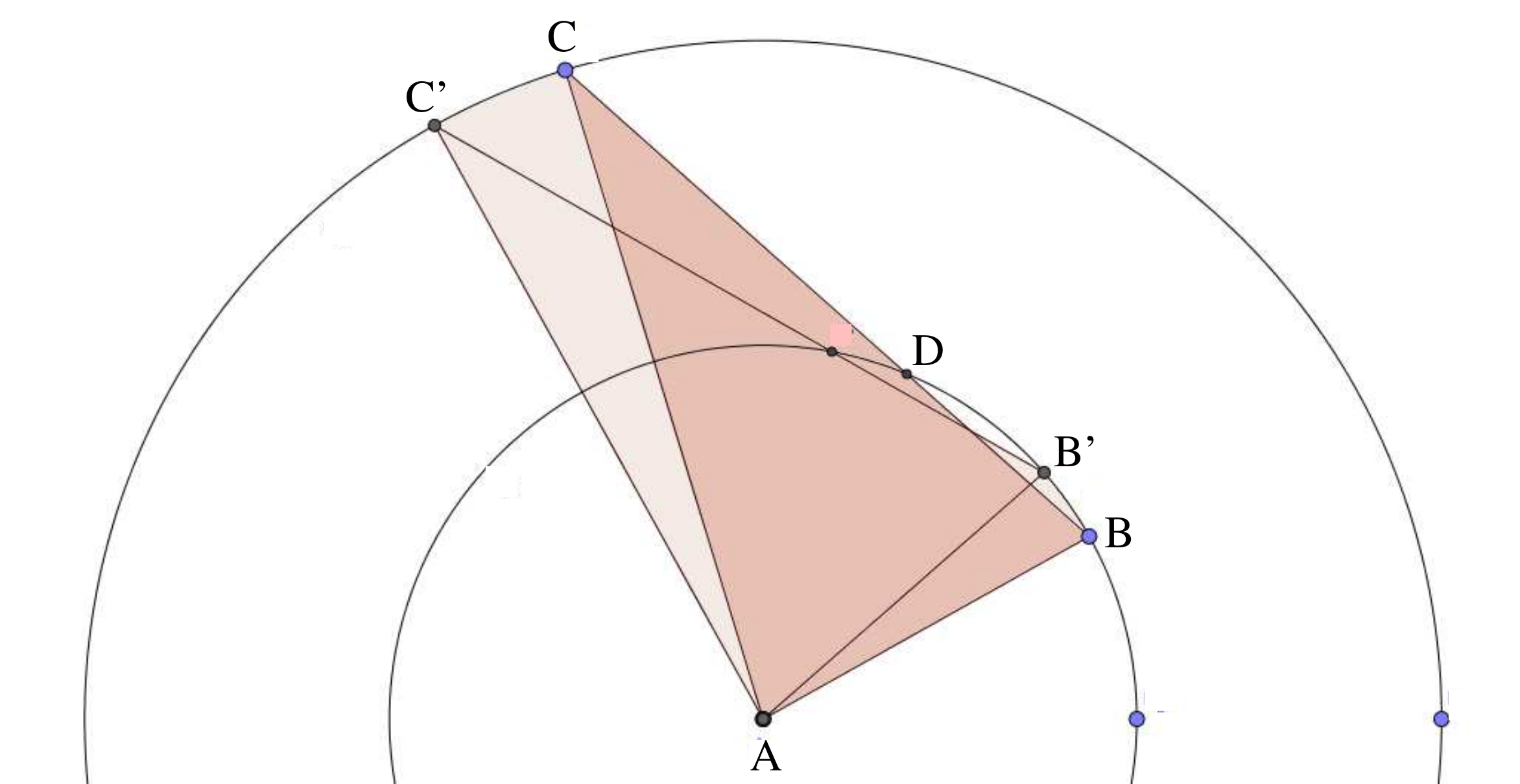}
	    	   \caption{
                   Proof of Lemma~\protect{\ref{lem:nice-tri-set}}: $T[0,\alpha]$ 
                   is not a truncated triangular set (TTS).}
	    	   \label{fig:not-nice-tri-set}
	    	  \end{center}
	    \end{figure} 	

} 

	\blem \label{lem:star-decompose}
{\em
        Let $R_0$ be a star-shaped polygonal region with $A$ as
        center.  If the boundary of $R_0$ is an $n$-gon, then we can
	decompose $R_0$ into an essentially disjoint\footnote{
	    A set $\set{A_1\dd A_k}$ where each $A_i\ib\RR^2$
	    is said to be \dt{essentially disjoint} if
	    the interiors of the $A_i$'s are pairwise disjoint.
	}
	union of at most
        $2n$ bounded triangular sets (i.e., at most $2n$ triangles)
        that are {\bf nice} and have $A$ as the apex.
}
	\elem
\remove{ 
\bpf 
        First, for each vertex $v$ of $R_0$ we add a segment
        connecting $A$ and $v$.  This decomposes $R_0$ into a disjoint
        union of $n$ triangles (since $R_0$ is star-shaped). Now
        consider each of the resulting triangle $T = [A,B,C]$ and let
        $A$ be the apex of $T$.  If $T$ is not nice, then both angles
        $b$ and $c$ (corresponding to vertices $B$ and $C$) are less
        than $90\degree$, and we can add a segment $[A,D]$ that is
        perpendicular to edge $[B,C]$ and intersects $[B,C]$ at $D$
        where $D$ is in the interior of $[B,C]$. This effectively
        decomposes $T$ into two {\em nice} triangles $[A,D,B]$ and
        $[A,D,C]$ with $A$ being the common apex. In this way, we can
        decompose $R_0$ into at most $2n$ nice triangles that have $A$
        as the apex.
\epf
} 

\subsection{Complex Predicates and T/R Subdivision Scheme} \label{se-star-soft}

For complex robots in general (not necessarily star-shaped), we can
exploit the remarkable \dt{decomposability} property of soft
predicates. More specifically, suppose $R_0=\cup_{j=1}^m T_j$ where each
$T_j$ is a triangle or other shapes and not necessarily pairwise
disjoint.  If we have soft predicates $\wtC_j(B)$ for each $T_j$
(where $B$ is a box), then we
immediately obtain a soft predicate for $R_0$ defined as follows:
	\begin{equation}
	    \wtC(B) = \clauses{\free &
	    		\textrm{ if each $\wtC_j(B)$ is $\free$}\\
		\stuck & \textrm{ if some $\wtC_j(B)$ is $\stuck$}\\
		\mixed & \textrm{ otherwise.}}
	\end{equation}
Let $\sigma>1$ and $\wtC$ be the soft version of an exact predicate $C$.
Recall
\cite{wang-chiang-yap:motion-planning:15,sss2}
that $\wtC$ is \dt{$\sigma$-effective} if for
all boxes $B$, if $C(B)=\free$ then $\wtC(B/\sigma)=\free$.

\bproDIY{A}{ \label{prop:soft-decompose}
\ \\ (1) $\wtC$ is a soft version of the exact classification predicate
for $R_0$.
\ \\    (2) Moreover, if each $\wtC_j$ is $\sigma$-effective, then
$\wtC$ is $\sigma$-effective.
}

We need $\sigma$-effectivity in soft predicates in order
to ensure resolution-exactness;
see \cite{wang-chiang-yap:motion-planning:15,sss2} where this proposition was
proved.
\ignore{
\bthm \label{prop:soft-decompose}
{\em
	The following is a soft predicate for $R_0$:

	Moreover, if each $\wtC_j$ is $\sigma$-effective, then
	then $\wtC$ is also $\sigma$-effective.
\ \\
}
	\ethm
}%
There are two important remarks.
First, this proposition is {\bf false}
if the $\wtC_j$ and $\wtC$ were exact predicates.
\noignore{
More precisely, suppose $C$ is the exact predicate for $R_0$
and $C_j$ is the exact predicate for each $T_j$.
It is true that if $C(B)=\free$ then $C_j(B)=\free$
for all $j$.  But if $C(B)=\stuck$, it does not follow that 
$C_j(B)=\stuck$ for some $j$.  
}%
Second, the predicates $\wtC_j(B)$ for
all the $T_j$'s must be based on a {\bf common coordinate system}.
As mentioned in Sec.~\ref{se-intro}, 
the soft predicate construction for a triangle robot in
\cite{wang-chiang-yap:motion-planning:15} does not work here.
A technical contribution of this paper is the design of soft predicates
$\wtC_j(B)$ for all the $T_j$'s that are based on a common coordinate
system. In the case of star-shaped robots, we apply
Lemma~\ref{lem:star-decompose} and use the apex $A$ as the origin of
this common coordinate system. 
%
%
Let $r_j$ be the length of the longer edge out of $A$ in $T_j$.  We
define $r_0$ as $r_0 = \max_j r_j$ (i.e., $r_0$ is the radius of the
circumcircle of $R_0$ centered at $A$).

{\bf T/R Splitting.}
The simplest splitting strategy is to split a box $B\ib\RR^d$ into
$2^d$ congruent subboxes.  In the worst case, to reduce all boxes
to size $<\eps$ requires time $\Omega(\log(1/\eps)^d)$; 
this complexity would not be practical for $d>3$.
In \cite{luo-chiang-lien-yap:link:14,yap-luo-hsu:thicklink:16}
we introduced an effective solution called {\em T/R splitting} which
can be adapted to configuration space\footnote{
    The configuration space of planar rigid robots is
	$SE(2)=\RR^2\times S^1$
	where $S^1$ is the unit circle representing angles $[0,2\pi)$.
}
$SE(2)$ in the current paper.
Write a box $B\ib SE(2)$ as a pair $(B^t,B^r)$ where
$B^t\ib\RR^2$ is the translational box and $B^r\ib S^1$ an
angular range $\Theta$.
%
%
We say box $B=(B^t,B^r)$ is \dt{$\vareps$-small} if
$B^t$ and $B^r$ are both $\vareps$-small;
the former means the width of $B^t$ is $\le \vareps$;
the latter means the angle (in radians) satisfies $|B^r|\le \vareps/r_0$.
Our splitting strategy is to only split $B^t$ (leaving $B^r=S^1$)
as long as $B^t$ is not $\vareps$-small. This is called a \dt{T-split},
and produces 4 children. Once $B^t$ is $\vareps$-small, we
do binary splits of $B^r$ (called \dt{R-split})
until $B^r$ is $\vareps$-small.  We discard $B$ when it is $\vareps$-small.
%
%
The following lemma (and proof) in~\cite{wang-chiang-yap:motion-planning:15}
can be carried over here:

\blem (\cite{wang-chiang-yap:motion-planning:15})
{\em
Assume $0< \vareps\le \pi/2$.
If $B = (B^t,B^r)$ is $\vareps$-small and $B^t$ is a square, 
%
%
then the Hausdorff distance between the footprints of $R_0$ at any two
configurations in $B$ is at most $(1+\sqrt{2})\vareps$.
%
}
\elem
\remove{ 
%
%
%
%
\bpf
This result uses the fact that if we rotate $R_0$ by $\theta$ about
the center of $R_0$, then the vertices of $R_1$ move by at most
$2r_0\sin (\theta/2)\le r_0\theta \leq \vareps$  
since $\sin\theta\le \theta$ for $\theta$ in the said range.  Also,
the translational distance between any two configurations in $B$ is at
most $\sqrt{2}\vareps$.
\epf
} 

{\bf Soft Predicates.}
Suppose we want to compute a soft predicate $\wtC(B)$ to classify
boxes $B$. Following the previous work
\cite{wang-chiang-yap:motion-planning:15,luo-chiang-lien-yap:link:14},
we reduce this to computing a feature set $\wtphi(B)\ib \Phi(\Omega)$.
The \dt{feature set} $\wtphi(B)$ of $B$ is defined as comprising those
features $f$ such that
	\beql{sepmb} 
	\Sep(m_B,f)\le r_B + r_0
	\eeql
\remove{
{\bf ***??? YJC: Here we use the ``global'' $r_0$ (see def.\ above) ---
Should we use an individual $r_j$ here (One feature set $\wtphi_j(B)$
per triangle $T_j$)? For now, we use the global $r_0$
and the global $\wtphi(B)$ for T-split boxes $B$, and separate
$\wtphi_j(B)$ for each $T_j$ for R-split boxes $B$.
==> This is OK.}\\
} 
where $m_B$ and $r_B$ are respectively the {\bf midpoint} and {\bf
  radius} of the translational box $B^t$ of $B = (B^t, B^r)$ (also
call them the {\bf midpoint} and {\bf radius} of $B$), and
%
%
    $\Sep(X,Y)\as \inf \{\|x-y\|:$ \\
%
%
$x \in X, y\in Y \}$ denotes the \dt{separation} of two
Euclidean sets $X,Y\ib\RR^2$.
We say that $B$ is \dt{empty} if $\wtphi(B)$ is empty but
$\wtphi(B_1)$ is not, where $B_1$ is the parent of $B$.  We may assume
the root is never empty.  If $B$ is empty, it is easy to decide
whether $B$ is \free\ or \stuck: since the feature set $\wtphi(B_1)$
is non-empty, we can find the $f_1\in\wtphi(B_1)$ such that
$\Sep(m_B,f_1)$ is minimized.  Then $\Sep(m_B,f_1)>r_B$, and by the
local property of features (see Feature-Based Approach in
Sec.~\ref{se-review}), we can decide if $m_B$ is inside ($B$ is \stuck)
or outside
$\Omega$ ($B$ is \free).

For a box $B$ where $B^r = S^1$, 
%
%
we maintain its feature set $\wtphi(B)$ as
above. But when $B^r \neq S^1$, 
%
%
we compute its feature set $\wtphi(B)$ as follows. Recall that we
decompose $R_0$ into a set of nice triangles $T_j$ with a common apex
$A$. For each $T_j$, consider the footprint of $T_j$ with $A$ at $m_B$
and rotating $T_j$ about $A$ from $\theta_1$ to $\theta_2$, where
$B^r = [\theta_1, \theta_2]$. By Lemma~\ref{lem:nice-tri-set} the resulting
swept area is a truncated triangular set (TTS); call it $TTS_j$.  We
define  
(cf.~\cite{wang-chiang-yap:motion-planning:15})
\ignore{
A \dt{shape} $S$ is be any closed simply-connected (possibly
unbounded) subset of $\RR^2$.  For $s > 0$,
%
%
}
for a 2D shape $S$ the $s$-\dt{expansion} of $S$, denoted by $(S)^s$, to be the Minkowski
sum of $S$ with the $Disc(s)$ of radius $s$ centered at the origin.
For a TTS, recall that $TTS = T \cap D$
where $T = H_1 \cap H_2 \cap H_3$ is an unbounded triangular
set (with each $H_i$ a half space) and $D$ is a disk
(\refFig{generalized-tri-b}). Note that $(TTS)^s$ is a proper subset
of $(H_1)^s \cap (H_2)^s \cap (H_3)^s \cap (D)^s$; a theorem in
the next section gives an exact representation of $(TTS)^s$.
We now specify the feature set $\wtphi(B)$:
for each $T_j$, let $\wtphi_j(B)$
comprise those features $f$ satisfying
%
%
$\Sep(m_B,f)\le r_B + r_j$ (replacing $r_0$ with $r_j$ in 
                            Eq.~(\ref{eq:sepmb})),  
%
%
such that $f$ also {\bf intersects the $r_B$-expansion of $TTS_j$}. We
can think of $\wtphi(B)$ as a collection of these $\wtphi_j(B)$'s,
each of which is used by the soft predicate $\wtC_j(B)$ so that we can
apply Proposition~A.
%

\sect[general]{General Complex Robots} \label{se-general}
%
When $R_0$ is a general polygon, not necessarily star-shaped,
we can still decompose $R_0$ into a set of triangles $T_j$
($j=1\dd m$), and consider
the rotation of these triangles relative to a fixed point $O$
(we may identify $O$ with the origin).
In this section, we define what it means for $T_j$ to be
``nice'' relative to a point $O$. 
If $O$ lies in the interior of $T_j$, we could decompose $T_j$
into at most $6$ nice pointed triangles at $O$, as in the previous section.
Henceforth, assume that $O$ does not lie in the interior of $T_j$.

%
\ssect[geom-general]{Basic Representation of Nicely Swept Sets}
%
	Let $T=[A,B,C]$ be any non-degenerate triangular region defined
	by the vertices $A, B, C$.   Let the origin $O$ be outside 
	the interior of $T$.  We define what
	it means for $T$ to be ``nice relative to $O$.''
	W.l.o.g., let $0\le \|A\|\le \|B\|\le\|C\|$ where $\|A\|$
	is the Euclidean norm.

	We say that $T$ is \dt{nice} if 
		the following three conditions hold:
		\beql{nice}
		\bang{A,B-A}\ge 0, \quad
		\bang{A,C-A}\ge 0, \quad
		\bang{B,C-B}\ge 0.
		\eeql
	Here $\bang{u,v}$ denotes the dot product of vectors $u,v$.

\begin{figure}[htb]  
		  \begin{center}
		    \begin{minipage}[t]{0.24\linewidth} 
                         \includegraphics[scale=0.20]{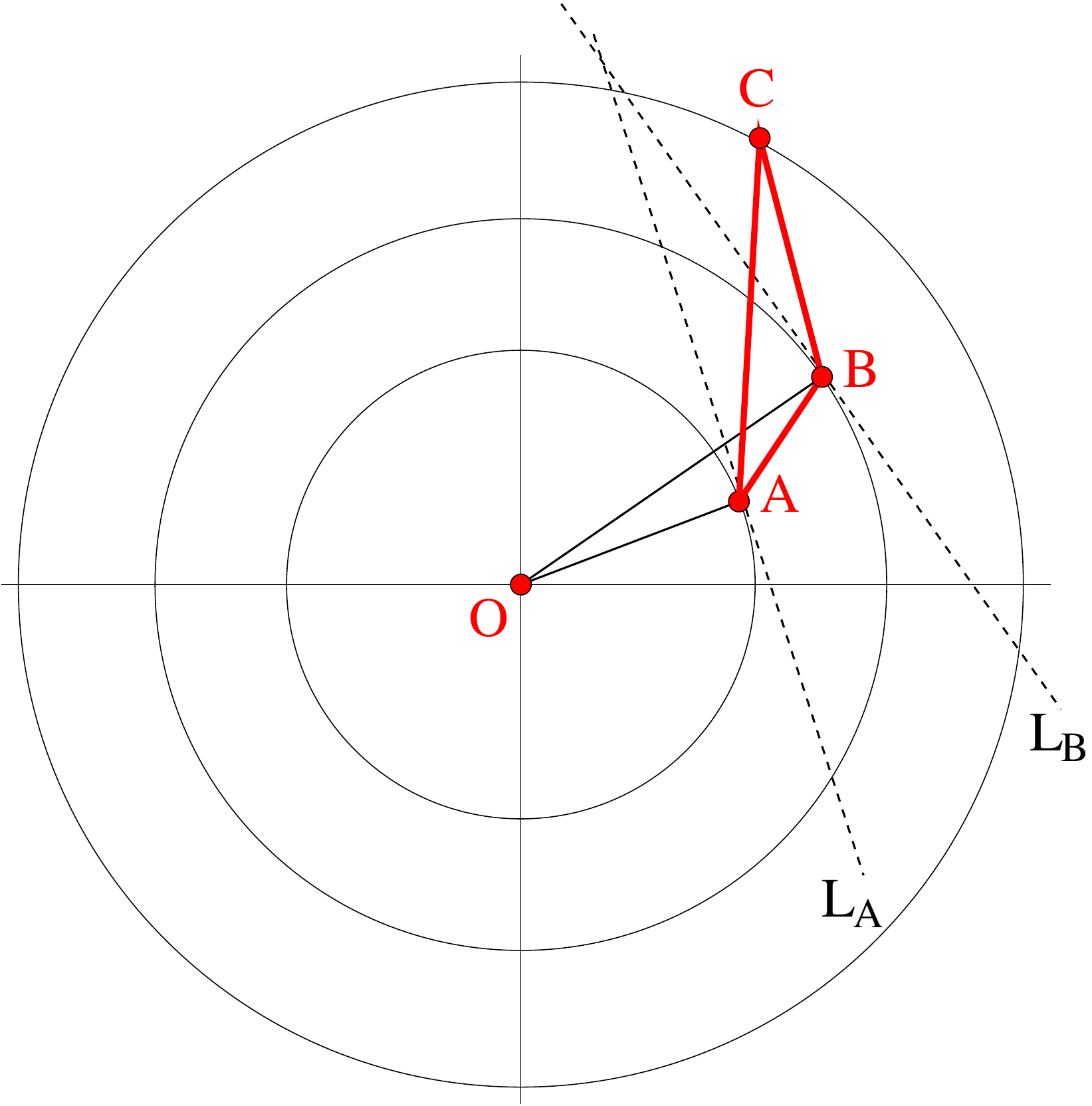}  
			 \caption{{\small Nice triangle $[A,B,C]$.}
                                 \label{fig:nice-tri-general}}
		    \end{minipage} \quad \quad \quad
	  	    \begin{minipage}[t]{0.26\linewidth}  
                        \includegraphics[scale=0.23]{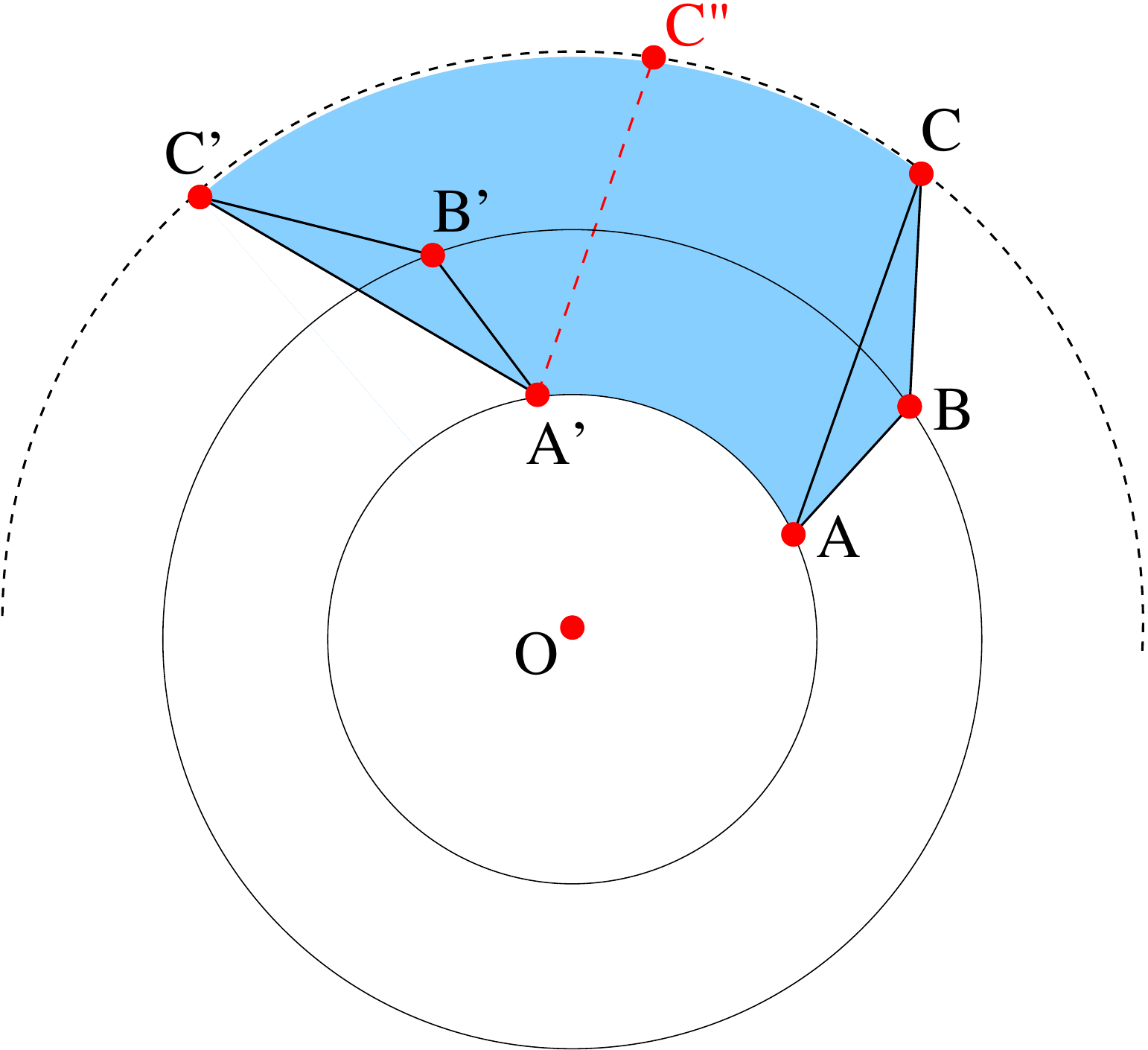}  
			\caption{{\small Nicely swept set (NSS, in blue)
				with $A, B, C$ in CCW order.} 
				 \label{fig:nicely-swept-set}}
	  	    \end{minipage} \quad \quad \quad
                    \begin{minipage}[t]{0.30\linewidth}  
                        \includegraphics[scale=0.23]{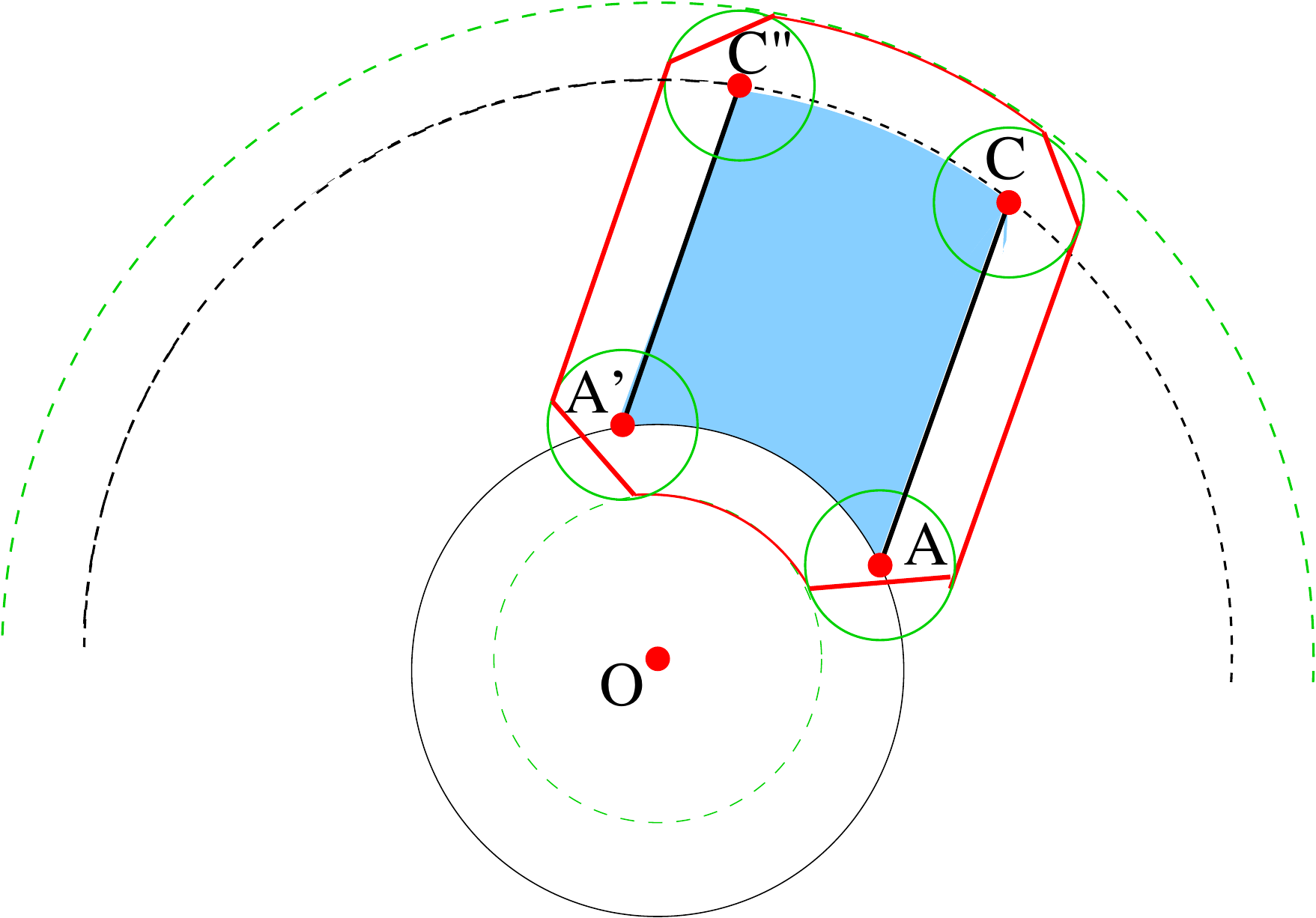} 
			\caption{{\small Expansion of  
                               %
			            $TruncStrip(A,C;A',C'')$ of Fig.~\protect{\ref{fig:nicely-swept-set}}
                                (in red).}
                                \label{fig:Minkowski-sum}}
                    \end{minipage}
	  	  \end{center}
  	\end{figure}

	A more geometric view of niceness is as follows
	(see~\refFig{nice-tri-general}).
	Draw three concentric circles centered at $O$ with
	radii $\|A\|,\|B\|,\|C\|$, respectively.  Two circles would
	coincide if their radii are equal, but we will see that
	the distinctness of the vertices and 
	niceness prevent such coincidences.
	Let $L_A$ be the line tangent to the circle of radius $\|A\|$
	and passing through the point $A$.
	Let $H_A$ denote the closed half-space bounded by $L_A$ and not containing $O$.
        The first condition in \refeq{nice}
	$\bang{A,B-A}\ge 0$ says that $B\in H_A$.  Similarly,
	the second condition says that $C\in H_A$.  
	Finally, the last condition says that $C\in H_B$ (where $H_B$
	is analogous to $H_A$).

	

%
%

%
\begin{figure}[htb] 
 \begin{center}
 \includegraphics[scale=0.23]{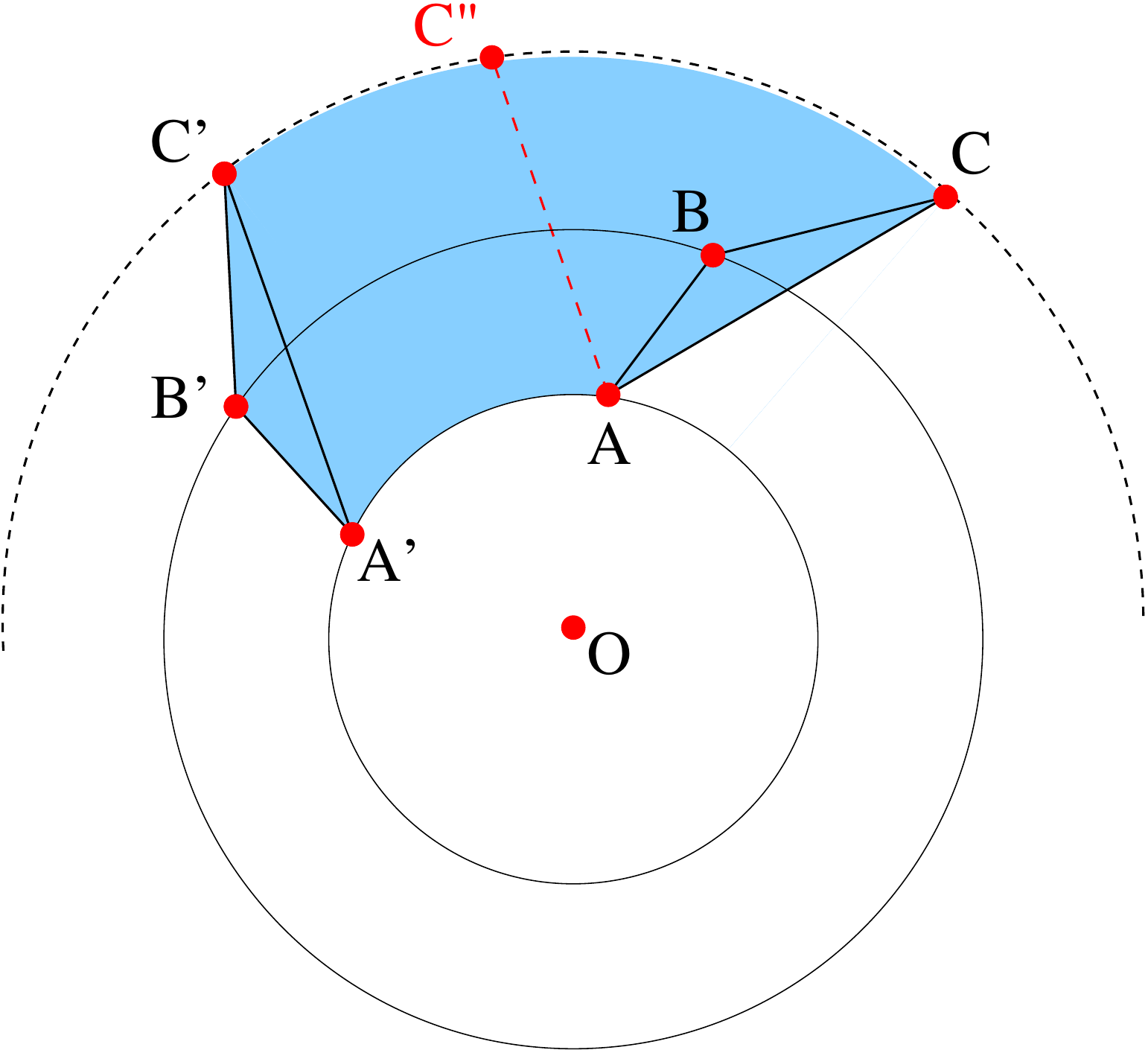}  
			\caption{{\small Nicely swept set (NSS, in blue)
				with $A, B, C$ in CW order.} 
				 \label{fig:nicely-swept-set-flip}}
 \end{center}
\end{figure}

\remove{
	Clearly, $T[\alpha,\beta]$ is contained in the annulus 
		$Disc(R)\setminus Disc(r)$
	where $r=\|A\| < \|C\| = R$ and
	$Disc(t)$ denotes the disc of radius $t$
	centered at $O$.
} 
	If $T$ is a nice triangle, then $T[\alpha,\beta]$ is
	called a \dt{nicely swept set (NSS)}.
	See ~\refFig{nicely-swept-set}, where $T[\alpha,\beta]$ is shaded in
	blue.  Let $T[\alpha]$ be the triangle $[A, B, C]$
        and $T[\beta]$ be $[A', B', C']$.
	W.l.o.g., assume\footnote{
		In case $A, B, C$ appear
		in clockwise (CW) order, the boundary of $T[\alpha,\beta]$
		can be similarly decomposed into two parts,
		comprising the swept segment $S[\alpha,\beta]$
		and the triangle $[A',B',C']$.
%
                See~\refFig{nicely-swept-set-flip}.
%
%
	}
	that $A, B, C$ appear in counter-clockwise (CCW) order
	as indicated
	in~\refFig{nicely-swept-set}.
	Then we can subdivide $T[\alpha,\beta]$ into two parts:
	the triangle $[A,B,C]$ and another part which
	we call a \dt{swept segment}. 

	\dt{Notation for Swept Segment:}
	if $S$ is the line segment $[A,C]$,
	then write $S[\alpha,\beta]$ for this swept segment.
        The boundary of $S[\alpha,\beta]$ is 
        decomposed into the
	following sequence of four curves given in clockwise (CW) order:
     (i) the arc $(A,A')$ centered at $O$ of radius $\|A\|$ from $A$ to $A'$,
     (ii) the segment $[A', C']$,
     (iii) the arc $(C',C)$ centered at $O$ of radius $\|C\|$ from $C'$ to $C$,
     (iv) the segment $[C, A]$.

	Our next goal is to consider $s$-expansion of the swept segment, i.e.,
		\beql{oplus}
		X = S[\alpha,\beta]\oplus Disc(s).
		\eeql
	Specifically, we want an easy way to detect the intersection
	between this expansion with any given feature (corner or edge).  
        To do so,
        we want to express $X$ as the union of ``basic shapes.''   
	A subset of $\RR^2$
	is a \dt{$0$-basic shape} if it is a half-space, a disc or complement
	of a disc.  We write $Disc(r)$ for the disc of radius $r$
	centered at $O$, and $Ann(r,r')$ for the annulus with inner
	radius $r$ and outer radius $r'$ centered at $O$.
	A shape $X$ is said to be \dt{$1$-basic} if it can be written as the
	finite intersection $X=\bigcap_{j=1}^k X_j$ where $X_j$'s
	are $0$-basic shapes.  
	The \dt{$1$-size} of $X$ is the minimum $k$ in such
	an intersection.  So polygons with $n$ sides
	have $1$-size of $n$.  Truncated
	triangular sets have $1$-size of $4$. We need some other
	$1$-basic shapes:
	\bitem
    \item \dt{Strips}:
	$Strip(a,b;a',b')$ is the region between the two parallel lines
	$\ol{a,b}$ and $\ol{a',b'}$.  
	Here $a,b,a',b'$ are distinct points.
    \item \dt{Truncated strips}:
	$TruncStrip(a,b;a',b')$ is the intersection of 
	$Strip(a,b;a',b')$ with an annulus; the boundary of 
	this shape is comprised of two line segments $[a,b]$ and $[a',b']$
	and two arcs $(a,a')$ and $(b,b')$ from the boundary of the annulus.
    \item \dt{Sectors}:
	$Sector(a,b,b')$ denotes any region bounded by a circular
	arc $(b,b')$ and two segments $[a,b]$ and $[a,b']$.
	\eitem

	Finally, a shape $X$ is said to be \dt{$2$-basic} if it can be
	written as a finite union of $1$-basic shapes,
	$X = \bigcup_{j=1}^m X_j$ where $X_j$'s are $1$-basic.
	We call $\set{X_1\dd X_m}$ a \dt{basic representation} of $X$.
	The \dt{$2$-size} of the representation is
	the sum of the $1$-sizes of $X_j$'s.  
	Thus, for any box $B^t\ib\RR^2$, the $s$-expansion of
	$B^t$ is a $2$-basic shape since it is
	the union of four discs and an octagon.
	We now consider the case where
	$X$ is the $s$-expansion of a swept segment $S[\alpha,\beta]$.
	We first decompose $S[\alpha,\beta]$ into
	two shapes as follows: suppose $C''$ lies on the circle
	of radius $\|C\|=\|C'\|$.  
%
%
Considering both cases of $A, B, C$ being in CCW and CW orders,
there are two possible representations:
	\\ (1) If $[A',C'']$ is parallel to $[A,C]$ and
	$[A',C'']\ib Ann(\|A\|,\|C\|)$, then we have 
		\beql{decomp1}
			S[\alpha,\beta]=
			Sector(A',C',C'') \cup TruncStrip(A,C;A',C'').
		\eeql
	\\ (2) If $[A,C'']$ is parallel to $[A',C']$ and
	$[A,C'']\ib Ann(\|A\|,\|C\|)$, then we have
		\beql{decomp2}
			S[\alpha,\beta]=
			Sector(A,C,C'') \cup TruncStrip(A,C'';A',C').  
		\eeql
	The swept segment in~\refFig{nicely-swept-set} supports the
        representation~\refeq{decomp1} but not~\refeq{decomp2},
%
        while the swept segment in~\refFig{nicely-swept-set-flip}
        supports the representation~\refeq{decomp2} but
        not~\refeq{decomp1}.  
        Note that they are symmetric cases, with
        $A, B, C$ in CCW order in~\refFig{nicely-swept-set} and in CW
        order in~\refFig{nicely-swept-set-flip}.
%
	Also, if the angular range of $[\alpha,\beta]$ is greater
	than $90$ degrees and the points $O,A,C$ are collinear,
	then both representations fail!
	We next show when at least one of the representations succeeds:

\blem
{\em
	Assume
	the width of the angular range $[\alpha,\beta]$ is at most $\pi/2$.
	Then swept segment $S[\alpha,\beta]$ can be
	decomposed into a sector and a truncated strip
	as in \refeq{decomp1} or \refeq{decomp2}.
}
\elem

Clearly, the $s$-expansion of a sector is $2$-basic.
%
%
This is also true for truncated strips (w.l.o.g., considering that in the
representation~\refeq{decomp1}):

\blem
{\em
	Let $X=TruncStrip(A,C; A',C'')$.
	There is a basic representation of $X\oplus D(s)$
	of the form $\set{D_1,D_2,D_3,D_4,X'}$ where $D_i$'s
	are discs and $X'$ is the intersection of a convex hexagon
	with an annulus.
}
\elem

Combining all these lemmas, we conclude:

\bthm
{\em
	Let $T[\alpha,\beta]$ be a 
	nicely swept set where $[\alpha,\beta]$ has width $\le \pi/2$.
	Then $T[\alpha,\beta]$ can be decomposed into 
	a triangle, a sector and a truncated strip.
	The $s$-expansion of $T[\alpha,\beta]$ has
	a basic representation which is the union of the
	$s$-expansions of the triangle, sector and truncated strip.
}
\ethm

The complexity of testing intersection of $2$-basic shapes
with any feature is proportional to its $2$-size, which is $O(1)$.
This theorem assures us
that the constants in ``$O(1)$'' is small.
%
%
%
Note that it is {\em not} correct to test if a line segment $L$
intersects a $1$-basic shape $X=\bigcap_{j=1}^k X_j$ by just testing
if $L$ intersects every $X_j$, since $L$ could intersect every $X_j$ but
not all in the same place(s) so that
  $L \cap X = \emptyset$.
%
%
%
Therefore, we need to maintain the {\em common intersections} between
$L$ and all $X_j$'s tested so far as we loop over all $X_j$'s; at the
end, $L$ intersects $X$ if and only if there is at least one non-empty
set of common intersections.  Since the complement of a disk is
non-convex, in general this process could result in many sets/segments
of common intersections to maintain. Fortunately, there is at most one
complement of a disk in our decomposition of an $NSS$.  Thus it is
enough to maintain just a {\em single} set/segment of the common
intersection of $L$ with all other $0$-basic shapes, and check with
the complement of a disk only at the end.
%

\remove{ 
Case (II):
	\bitem
    \item the arc centered at $O$ of radius $r$ from $A$ to $A'$  
    \item the segment $[A', B']$                                  
    \item the segment $[B', C']$                                  
    \item the arc centered at $O$ of radius $R$ from $C'$ to $C$  
    \item the segment $[C, A]$.                                   
	\eitem
}

\ignore{
	Consider the triangular set $S=H_{A,B}\cap H_{B,C} \cap H_{C',A'}$,
	where the notation $H_{A,B}$ refers to the half space to the left of
	the line passing through $A$ and $B$, with orientation from $A$ to
	$B$. Also, let $Disc(t)$ denote the disc of radius $t$ centered at
	$O$. Then the nicely swept set (NSS) $T[\alpha,\beta]$ just defined
	(see~\refFig{nicely-swept-set}) is exactly $(S\cap Disc(\|C\|))
	\setminus Disc(\|A\|)$.
}%

%

\remove{ 
	\bpf
W.l.o.g., assume that $A,B,C$ are in a CCW order (the other case is
symmetric).  Suppose $T$ is nice.  We claim that if $\beta - \alpha >
0$ is small enough, then $T[\alpha,\beta]$ is equal to $(S\cap
Disc(\|C\|)) \setminus Disc(\|A\|)$.
To see this, 
look at~\refFig{nice-tri-general} and the conditions of niceness
in~(\ref{eq:nice}): the first two conditions makes sure that $B \in
H_A$ and $C \in H_A$, i.e., the edge $[A, B]$ (resp.\ $[A,C]$) does
not cut through $Circle A$ (denoting the circle centered at $O$ and
passing through $A$) and thus $T[\alpha,\beta]$ is outside
$Disc(\|A\|)$. Similarly, the third condition makes sure that $C \in
H_B$, i.e., the edge $[B, C]$ does not cut through $Circle
B$. Therefore $[B, C]$ is always above $Circle B$ during sweeping, and
the shape of $T[\alpha,\beta]$ is $(S\cap Disc(\|C\|)) \setminus
Disc(\|A\|)$.
%
%

	Conversely, suppose $T$ is not nice.  Say the first condition
        $\bang{A,B-A}\ge 0$ is violated.  This means that the edge
        $[A,B]$ intersects $Circle A$ at another point $D\in [A,B]$
        (see~\refFig{Not-nicely-swept2}(a)). Consider rotating $[A, B,
          C]$ to $[A', B', C']$ such that $A'$ is between $A$ and
        $D$. Clearly the corresponding swept set $T[\alpha,\beta]$ has
        some portion inside $Disc(\|A\|)$ and thus it cannot be
        $(S\cap Disc(\|C\|)) \setminus Disc(\|A\|)$.  Violating the
        second condition $\bang{A,C-A}\ge 0$ is similar.  Now consider
        violating the third condition $\bang{B,C-B}\ge 0$.  This means
        that the edge $[B,C]$ intersects $Circle B$ at another point
        $D\in [B,C]$ (see~\refFig{Not-nicely-swept2}(b)). Consider
        sweeping from $[A, B, C]$ to $[A', B', C']$ where $B'$ is
        between $B$ and $D$. Then $[B,C]$ and $[B',C']$ intersect at a
        point $E$ below $Circle B$ and we see that the boundary
        portion $[B', E, D]$ of the swept set $T[\alpha,\beta]$ is
        {\bf non-convex}.  Note that a nicely swept set (NSS), before
        subtracting $Disc(\|A\|)$, i.e., $S\cap Disc(\|C\|)$, is {\bf
          convex}, but $T[\alpha,\beta]$ is non-convex even before
        subtracting $Disc(\|A\|)$ (due to $[B', E, D]$), so
        $T[\alpha,\beta]$ is not an NSS.
\epf

	    \begin{figure}[htb]
	    	  \begin{center}
		   \scalebox{0.28}{
	    	     \input{./figs/Not-nicely-swept2.pdf_t}}
	    	   \caption{Proof of Theorem~\protect{\ref{thm-NSS}}:
  The area swept from triangle $[A, B, C]$ to $[A', B', C']$ is not a
  nicely swept set (NSS).}
	    	   \label{fig:Not-nicely-swept2}
	    	  \end{center}
	    \end{figure} 	

} 

\ssect[partition]{Partitioning an $n$-gon into Nice Triangles}
%
	Suppose $P$ is an $n$-gon.  We can partition it into $n-2$
        triangles.  W.l.o.g., there is at most one triangle that
        contains the origin $O$.  We can split that triangle into at
        most 6 nice triangles, using our technique for star-shaped
        polygons (Lemma~\ref{lem:star-decompose}).

	\blem \label{lem:4-nice-triangles}
{\em
	If $T$ is an arbitrary triangle and $O$ is exterior to $T$,
	then we can partition $T$ into at most $4$ nice triangles.
}
	\elem

	The number $4$ in this lemma is the best possible:
	if $T$ is a triangle with circumcenter $O$,
	then any partition of $T$ into nice triangles
	would have at least $4$ triangles because we need
	to introduce vertices in the middle of each side of $T$.
\remove{ 
    \bpf 
        We refer to~\refFig{4-nice-triangles}. Note that segments
        $[O,D], [O,E]$ and $[O,F]$ are perpendicular to segments
        $[A,B], [A,C]$ and $[B,C]$, respectively. It can be easily
        verified that each of the 4 triangles $[D,A,E], [D,E,C],
        [D,C,F]$ and $[D,F,B]$ is a nice triangle.
    \epf

	    	\begin{figure}[htb]
	    	  \begin{center}
		   \scalebox{0.3}{
	    	     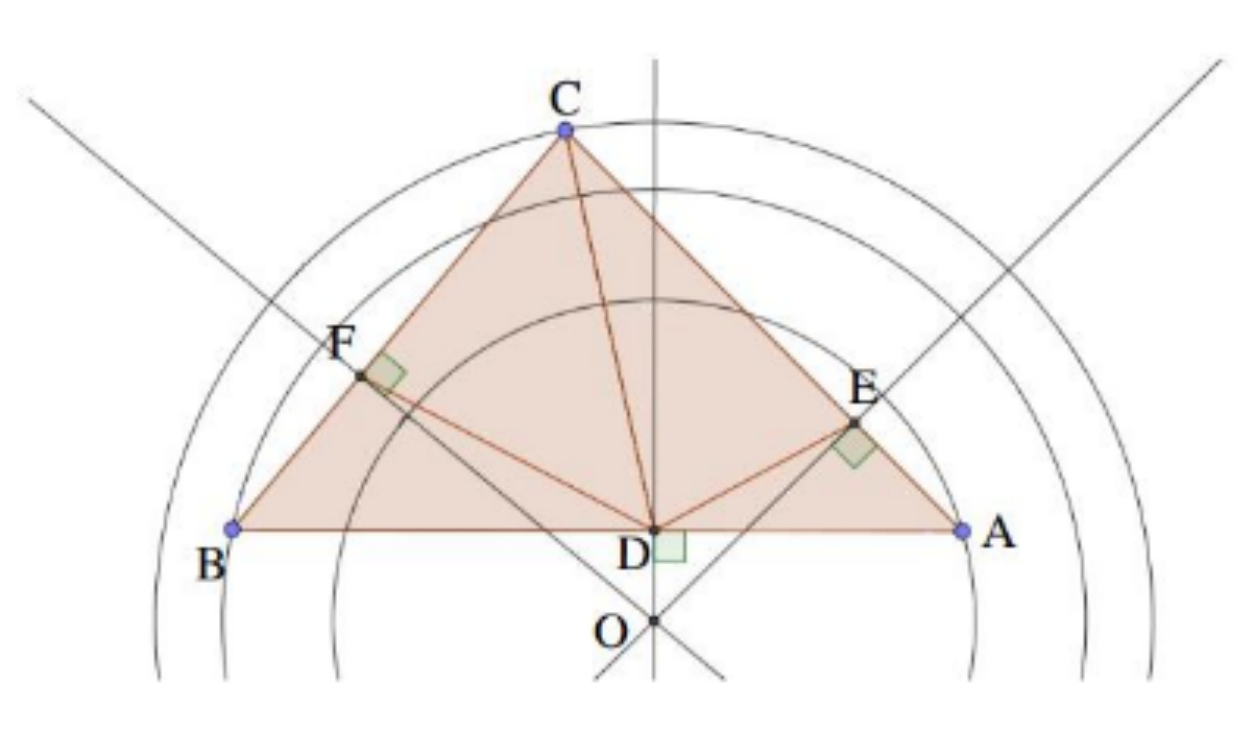}
	    	   \caption{Proof of Lemma~\protect{\ref{lem:4-nice-triangles}}: 
                    A triangle $T = [A,B,C]$ with the origin $O$ in the exterior
                    can be decomposed into 4 nice triangles.}
	    	   \label{fig:4-nice-triangles}
	    	  \end{center}
	    	\end{figure} 	
} 

	\bthm  \label{thm-general-n-gon-partition}
{\em
	Let $P$ be an $n$-gon.
	\\(i)
	Given any triangulation of $P$
	into $n-2$ triangles, we can refine the triangulation 
	into a triangulation with $\le 4n-6$ nice triangles.
	\\(ii)
	This bound is tight in this sense: for every $n\ge 3$,
	there is a triangulation of $P$ whose refinement has size $4n-6$.
\ \\
}
	\ethm
\remove{ 
	\bpf
	(i) In the given triangulation of $P$, we might
	have a triangle containing $O$.  This triangle can be
	triangulated into at most $6$ nice triangles (Lemma~\ref{lem:star-decompose}).
	By Lemma~\ref{lem:4-nice-triangles},
        the remaining $n-3$ triangles can be 
	refined into $4(n-3)$ nice triangles. 
        The final count is $6+4(n-3)=4n-6$. 
	\\ (ii) We construct an $n$-gon whose vertices are all
	on the unit circle.  Note that all such vertices
	are of the form $e^{\ii\theta}$.
	For the first triangle $T_0$, pick the vertices 
	$1, e^{\ii 2\pi/3}, e^{\ii 4\pi/3}$.
%
%
        Call these vertices $u_0, u_1, u_2$. Choose the origin $O$ inside $T_0$
        so that for each triangle $[O, u_i, u_j], i \neq j \in \set{0, 1, 2}$
        we have both $\angle Ou_iu_j < 90\degree$ and $\angle Ou_ju_i < 90\degree$.
        Therefore each triangle $[O, u_i, u_j]$ must be split into $2$ nice triangles
        and overall $T_0$ must be split into $6$ nice triangles.
%
%
%
	If $n=3$, our result is verified.  If $n>3$,
	we must add $n-3$ additional vertices.
	Define the vertex
	$v_k \as e^{\ii \frac{2k\pi}{3(n-2)}}$, $k=0,1\dd n-2$
%
%
	(note that $v_0=1$ and $v_{n-2}=e^{\ii 2\pi/3}$ have
	previously been chosen).  Thus we have added $v_1\dd v_{n-3}$
	new vertices.
	Any triangle $[v_k, v_\ell, v_m]$ must be split into $4$ nice triangles.
	This proves our claim.
	\epf

} 

%
\remove{ 
\vspace*{-2em}
{\bf Remark.} 
                Part (ii) of this theorem
shows that for any $n\ge 3$, there is an $n$-gon with an origin $O$
such that {\em every} nice partition of the $n$-gon decomposes it into
$4n-6$ nice triangles.
%
%
} 

\ssect[general-soft]{Soft Predicates and T/R Subdivision Scheme}
%
We can now follow the same paradigm as for star-shaped robots in
Sec.~\ref{se-star-soft}. We first apply
Theorem~\ref{thm-general-n-gon-partition}(i) to partition the robot
$R_0$ into a set of nice triangles, $R_0=\cup_j T_j$, where all
$T_j$'s share a common origin $O$, and we will use the soft predicates
developed for $T_j$ and apply 
Proposition~A. 
%
%
The origin $O$ plays a similar role as the apex in
Sec.~\ref{se-star-soft}.  The T/R splitting scheme is exactly the
same: we first perform T-splits, splitting only the translational
boxes until they are $\vareps$-small, and then we perform R-splits,
splitting only the rotational boxes until they are
$\vareps$-small. Essentially the top part of the subdivision tree is a
quad-tree, and the bottom parts are binary subtrees (see
Sec.~\ref{se-star-soft}).

The feature set for a subdivision box $B$ where we perform T-splits is
the same as before; the only difference is that now for a box $B$
where we perform R-splits, we use a new feature set $\wtphi_j(B)$ for
each nice triangle $T_j$ where $O$ is not at its vertex (there are at
most 6 nice triangles with $O$ at a vertex/apex; see
Theorem~\ref{thm-general-n-gon-partition}(i)).  Suppose $T_j =
[a,b,c]$ with $0\le \|a\|\le \|b\|\le\|c\|$. Let $r_j = \|c\|$. Also,
suppose the angle range of box $B = (B^t, B^r)$ is $B^r = [\theta_1,
  \theta_2]$.  Recall the footprint of $T_j[\theta_1,\theta_2]$
is a nicely swept set (NSS); denote it $NSS_j$. Then the new feature
set $\wtphi_j(B)$ for $T_j$ comprises those $f$ where
$\Sep(m_B,f)\le r_B + r_j$ and $f$ also {\bf intersects the
  $r_B$-expansion of $NSS_j$}
(where $m_B$ and $r_B$ are the midpoint and radius of $B$).
%

\sect[results]{Experimental Results} \label{se-results}
%
%
\begin{table}[tbh]
\centering
\caption{Robot Statistics.}
\label{tab:rob-stats}
    {\small 
\begin{tabular}{l|ll}
\hline
Robot       & $m$ (\# sides) &  $t$ (\# triangles) \\
\hline
L-shaped    & 6              &  4    \\
snowflake   & 18             & 24    \\
S-shaped    & 12             & 26    \\
3-legged    & 14             & 20    \\
C-shaped    & 18             & 22    \\
\hline
\end{tabular}
    }
\end{table}
%
%
%
%
\begin{table}[tbh]
\centering
\caption{Running Our Planner 
 (R: radius of the robot's circumcircle around its rotation center; P?: path found? (Yes/No); Time
is in s; S-shaped*: thin version).} 
%
%
\label{tab:exp-result}           
     {\small 
\begin{tabular}{lllllllll}
\hline
\multicolumn{1}{l|}{Exp\# } & Robot       & Envir. & R & $\epsilon$ & $\alpha$                 & \multicolumn{1}{l|}{$\beta$}                  & P?   & Time \\\hline\hline

\multicolumn{1}{l|}{0}   & L-shaped    & gateway    & 50     & 2          & (18, 98, 340$\degree$)   & \multicolumn{1}{l|}{(458,119,270$\degree$)}   & Yes    & 10.106   \\
\multicolumn{1}{l|}{1}   & L-shaped    & gateway    & 50     & 4          & (18, 98, 340$\degree$)   & \multicolumn{1}{l|}{(458,119,270$\degree$)}   & No     & 8.431   \\
\hline
\multicolumn{1}{l|}{2}   & snowflake   & sparks     & 56     & 2          & (108, 136, 0$\degree$)   & \multicolumn{1}{l|}{(358, 155, 0$\degree$)}   & Yes     & 17.846  \\
\multicolumn{1}{l|}{3}   & snowflake   & sparks     & 56     & 2          & (108, 136, 0$\degree$)   & \multicolumn{1}{l|}{(358, 155, 180$\degree$)} & Yes     & 3.370   \\
\hline
\multicolumn{1}{l|}{4}   & S-shaped    & sparks     & 74     & 4          & (132, 80, 90$\degree$)   & \multicolumn{1}{l|}{(333, 205, 90$\degree$)}  & Yes     & 34.284   \\
\multicolumn{1}{l|}{5}   & S-shaped    & sparks     & 74     & 4          & (132, 80, 90$\degree$)   & \multicolumn{1}{l|}{(333, 205, 60$\degree$)}  & No      & 57.371   \\
\hline
\multicolumn{1}{l|}{6}   & 3-legged    & sparks     & 70     & 2          & (108, 136, 0$\degree$)   & \multicolumn{1}{l|}{(368, 155, 0$\degree$)}   & Yes      & 41.745   \\
\hline
%
\multicolumn{1}{l|}{7}   & L-shaped    & corridor   & 68     & 2          & (75, 420,  0$\degree$)   & \multicolumn{1}{l|}{(370, 420, 0$\degree$)}  & Yes      & 4.012   \\
\multicolumn{1}{l|}{8}   & L-shaped    & corridor   & 68     & 3          & (75, 420,  0$\degree$)   & \multicolumn{1}{l|}{(370, 420, 0$\degree$)}  & Yes      & 1.926   \\
\multicolumn{1}{l|}{9}   & L-shaped    & corridor   & 68     & 5          & (75, 420,  0$\degree$)   & \multicolumn{1}{l|}{(370, 420, 0$\degree$)}  & Yes      & 2.684  \\
\multicolumn{1}{l|}{10}   & L-shaped    & corridor-L & 68     & 5          & (75, 420,  0$\degree$)   & \multicolumn{1}{l|}{(370, 420, 0$\degree$)}  & No      & 2.908  \\
\multicolumn{1}{l|}{11}   & L-shaped    & corridor-L & 68     & 3          & (75, 420,  0$\degree$)   & \multicolumn{1}{l|}{(370, 420, 0$\degree$)}  & Yes      & 2.255  \\
\hline
\multicolumn{1}{l|}{12}   & C-shaped    & corridor-S & 80     & 4          & (80, 450,  0$\degree$)   & \multicolumn{1}{l|}{(380, 450, 0$\degree$)}  & Yes      & 26.200  \\
\hline
%
%
\multicolumn{1}{l|}{13}   & S-shaped       & maze       & 38     & 2          & (38, 38,  0$\degree$)   & \multicolumn{1}{l|}{(474, 474, 90$\degree$)}  & No       & 90.097  \\
\multicolumn{1}{l|}{14}   & S-shaped* & maze       & 38     & 2          & (38, 38,  0$\degree$)   & \multicolumn{1}{l|}{(474, 474, 90$\degree$)}  & Yes      & 79.518  \\  
\hline

\end{tabular}
    }
\end{table}
%
%
%
%
\begin{table}[tbh]
\centering
\caption{Comparing with OMPL (``\#'': Exp\#; ``Time/P?'': our run time (in s)/path found? (Y/N). 
Each OMPL method: Average Time (in s)/Standard Deviation/Success Rate, over 10 runs).
}
\label{tab:exp-ompl}
    {\small  
\begin{tabular}{l||l|llll}
\hline

\#  &    Time/P?    &          PRM      &          RRT         &         EST            &   KPIECE \\
\hline\hline
0   &     10.106/Y     &  {\bf 4.18}/2.53/1      &  42.13/38.49/1       &    76.22/110.44/0.9    &    300/0/0  \\
2   &     17.846/Y     &  {\bf 9.22}/6.82/1      &  210.41/144.25/0.3   &   271.75/89.31/0.1     &    240.00/126.47/0.2 \\
3   &      {\bf 3.370}/Y     &  300/0/0          &  300/0/0             &   300/0/0              &    300/0/0          \\
4   &     34.284/Y     &  {\bf 5.93}/7.20/1      &  217.33/134.53/0.3   &   300/0/0              &    300/0/0          \\
5   &     {\bf 57.371}/N     &  300/0/0          &  300/0/0             &   300/0/0              &    300/0/0          \\
6   &     41.745/Y     &  {\bf 2.72}/4.89/1      &  154.22/141.77/0.5   &   104.32/78.10/0.7     &    3.16/4.28/1   \\
%
%
8   &      1.926/Y     &  0.63/0.55/1      &  300/0/0             &    3.02/4.71/1         &    {\bf 0.41}/0.28/1     \\
11  &      2.255/Y     &  {\bf 1.49}/0.84/1      &  300/0/0             &   241.24/124.88/0.2    &    1.58/1.47/1    \\
12  &     26.200/Y     &  {\bf 3.16}/4.21/1      &  300/0/0             &   172.506/120.38/0.7   &   93.88/88.03/0.8   \\
13  &     {\bf 90.097}/N  &     300/0/0          &  300/0/0             &   300/0/0              &    300/0/0          \\
14  &     79.518/Y        &     300/0/0          &  236.72/106.44/0.3   &   300/0/0              &    {\bf 39.81}/91.57/0.9 \\
\hline
\end{tabular}
   }
\end{table}
%
%
%
%
%
\FigEPS{envir-6}{0.07186}{Six Environments in our experiments.
\label{fig:six-env}} 
We have implemented our approaches in {\tt C/C++} with
{\ttt Qt} GUI platform.
The software and data sets 
are freely available from 
the web site for our open-source 
\corelib~\cite{core-download-link}.
%
%
%
%
%
All experiments are reproducible as targets of Makefiles in 
\corelib.
Our experiments are on a PC with one 3.4GHz Intel Quad Core i7-2600 
CPU, 16GB RAM, nVidia GeForce GTX 570 graphics 
%
%
and Linux Ubuntu 16.04 OS.
The results are summarized in \refTab{exp-result} and
\refTab{exp-ompl}. \refTab{exp-result} is concerned only with
the behavior of our complex robots; \refTab{exp-ompl} gives comparisons
with the open-source OMPL library~\cite{OMPL}.
%
%
The robots are as shown in~\refFig{robot};
their statistics are given in~\refTab{rob-stats}.
%

\ignore{
There are two sets
of experiments, for two complex robots: an $L$-shape robot and the
snow flake robot in \refFig{robot}(d).  For each robot, we run it in
one of the four environments in \refFig{envir-new} or in the parking
environment in \refFig{gui}.  In \refTab{exp-result}, radius is a
robot parameter. More specifically, it is the distance from the origin
to the furthest vertex of a robot. The resolution exactness parameter
is $\epsilon$, and $\alpha$ and $\beta$ are the start and goal
configurations. The column \#Box gives the total number of
configuration boxes that are created in the subdivision search to find
a path (or to conclude \nopath).
}

We select some interesting experiments to 
  analyze
%
%
%
characteristic behavior of our planner.  Please see
\refTab{exp-result} and the video
%
(\myHrefxx{https://cs.nyu.edu/exact/gallery/complex/complex-robot-demo.mp4}).
In Exp0-1, we show how the parameter $\epsilon$ affects the result.
With a narrow gateway,
when we change $\eps$ from $2$ to $4$, the output changes
from a path to \nopath~for the same configuration.
In Exp2-3, we observe how the snowflake robot
rotates and maneuvers to get from the start
to two different goals. 
For Exp4-5, the difference is in the angles of the goal
configuration; 
in Exp5 this is designed to be an isolated configuration and the planner outputs
\nopath~ as desired. 
%
%
Exp6 shows how the robot squeezes among the obstacles to move its
complex shape through the environment.
%
%
%
Exp7-9 use the same L-shaped robot, $\alpha, \beta$ configurations and
the environment; only $\eps$ varies.
%
%
The planner can find three totally different paths.
When $\eps$ is small (Exp7), the path is very carefully adjusted to
move the robot around the obstacles. When $\eps$ is larger (Exp8), the
planner finds an upper path with a higher clearance.  When $\eps$ is
even larger (Exp9), the planner chooses a very safe but much longer
path at the bottom.
Note that using a larger $\eps$ usually makes the search faster, since
we stop splitting boxes smaller than $\eps$, but a longer path can
make the search slower.
%
%
In Exp10-11, we modify the environment of Exp7-9 by putting a large
obstacle at the bottom, which forces the robot to find a path at the
top.
Exp12 uses an environment similar to those in Exp7-11 but with much
smaller scattered obstacles.  It is designed for the C-shaped robot,
which can rotate while having an obstacle in its pocket.
%
%
Exp13-14 use a challenging environment where the small scattered
obstacles force the S-shaped robot to rotate around and only the ``thin''
version (Exp14, also in Fig.~\ref{fig:six-env} ``maze'') can squeeze
through.

In Table~\ref{tab:exp-ompl} we compare our planner with several
sampling algorithms in OMPL: PRM, RRT, EST, and
KPIECE. These experiments 
are correlated to those
in \refTab{exp-result} (see the Exp \#).  Each OMPL planner is run 10
times with a time limit 300 seconds (default), where all
planner-specific parameters use the OMPL default values. We see that
for OMPL planners there are often unsuccessful runs and they have to
time out even when there is a path.  On the other hand, our algorithm
consistently solves the problems in a reasonable amount of time, often
much faster than the OMPL planners, in addition to being able to
report \nopath. 


\sect[conclude]{Conclusions}  \label{se-conclude}

Although the study of rigorous algorithms for motion
planning has been around for over 40 years, 
there has always been a gap
between such theoretical algorithms and the practical methods.
Our introduction
of resolution-exactness and soft predicates on
the theoretical front, together with matching implementations,
closes this gap.  Moreover, it eliminated the ``narrow passage''
problem that plagued the sampling approaches.
The present paper extends our approach to challenging
planning problems for which no exact algorithms exist.
%

What are the current limitations of our work?  We implement
everything in machine precision (the practice
in this field).  
%
%
  But it can be easily modified to
achieve the theoretical guarantees of resolution-exactness if we use
arbitrary precision BigFloats number types.
%

We pose two open problems: One is to find an optimal 
decomposition of $m$-gons into nice triangles (currently, we
simply give an upper bound).  Such decompositions will
have impact for practical complex robots.
Second, we would like to develop similar decomposability
of soft predicates for complex rigid robots in $\RR^3$.



%
\bibliographystyle{abbrv} 
%
\bibliography{test,st,yap,exact,geo,alge,math,com,rob,cad,algo,visual,gis,quantum,mesh,tnt,fluid}
%
%

%
%
\section*{APPENDIX: Proofs}

%

\noindent {\bf Lemma 1.}	
{\em
		Let $T$ be a pointed triangular set.
		Then $T$ is nice iff
		for all $\alpha\in S^1$ ($0<\alpha<\pi-a$),
		the footprints
		$T[0,\alpha]$ and $T[-\alpha,0]$
		are truncated triangular sets (TTS).
}

\bpf
	If $T$ is nice, $T[0,\alpha]$ and $T[-\alpha,0]$ are 
        truncated triangular sets (TTS); this is easily seen in 
	\refFig{generalized-tri-b}(c).

        Conversely, if $T$ is not nice,
        let us assume that $\|A-B\|\le \|A-C\|$ (e.g.,
        \refFig{generalized-tri-b}(d)).  We claim that for
        sufficiently small $\alpha>0$, either $T[0,\alpha]$ or
        $T[-\alpha,0]$ is not a TTS.
	Assume (w.l.o.g.) that $A,B,C$ are in CCW order;
	we show that $T[0,\alpha]$ is not a TTS.
%
%
%

	If $T$ is not nice, then $b<90\degree$. Let $B-C$ intersects the
	$Circle B$ (the circle centered at $A$ that passes through $B$) 
	at $D$. Let $\alpha_{max} = \angle BAD
	= 180\degree-2b = 2(90\degree-b)$, 
since $b = \angle ABD = \angle ADB$.
        Note that a TTS is a {\bf convex} set as it is the intersection
        of three half-spaces and one disc; all of them are convex and thus
        the intersection is also convex.
	However, for any $\alpha < \alpha_{max}$, $T[0,\alpha]$ is not a TTS
	since $B-C$ will intersect $B'-C'$ inside $Circle B$ (see
	\refFig{not-nice-tri-set}) --- 
%
%
%
  this makes $T[0,\alpha]$ {\bf non-convex} and thus it is not a TTS.
%
%
%
	\epf


	    \begin{figure}[htb]
	    	  \begin{center}
		   \scalebox{0.3}{
	    	     \input{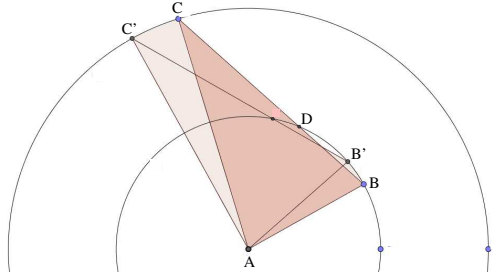}}
	    	   \caption{
                   Proof of Lemma~\protect{\ref{lem:nice-tri-set}}: $T[0,\alpha]$ 
                   is not a truncated triangular set (TTS).}
	    	   \label{fig:not-nice-tri-set}
	    	  \end{center}
	    \end{figure} 	

\noindent {\bf Lemma 2.}
{\em
        Let $R_0$ be a star-shaped polygonal region with $A$ as
        center.  If the boundary of $R_0$ is an $n$-gon, then we can
        decompose $R_0$ into an essentially disjoint union of at most
        $2n$ bounded triangular sets (i.e., at most $2n$ triangles)
        that are {\bf nice} and have $A$ as the apex.
}

\bpf 
        First, for each vertex $v$ of $R_0$ we add a segment
        connecting $A$ and $v$.  This decomposes $R_0$ into a disjoint
        union of $n$ triangles (since $R_0$ is star-shaped). Now
        consider each of the resulting triangle $T = [A,B,C]$ and let
        $A$ be the apex of $T$.  If $T$ is not nice, then both angles
        $b$ and $c$ (corresponding to vertices $B$ and $C$) are less
        than $90\degree$, and we can add a segment $[A,D]$ that is
        perpendicular to edge $[B,C]$ and intersects $[B,C]$ at $D$
        where $D$ is in the interior of $[B,C]$. This effectively
        decomposes $T$ into two {\em nice} triangles $[A,D,B]$ and
        $[A,D,C]$ with $A$ being the common apex. In this way, we can
        decompose $R_0$ into at most $2n$ nice triangles that have $A$
        as the apex.
\epf

\noindent {\bf Lemma 4.} {\em
	Assume
	the width of the angular range $[\alpha,\beta]$ is at most $\pi/2$.
	Then swept segment $S[\alpha,\beta]$ can be
	decomposed into a sector and a truncated strip
	as in \refeq{decomp1} or \refeq{decomp2}
} 

\bpf
Our goal is to choose the point $C''$ so that either
\refeq{decomp1} or \refeq{decomp2} holds.
Let the swept segment be $S[\alpha,\beta]$, with $S[\alpha]=[A,C]$
and $S[\beta]=[A',C']$.  Let $D$ (resp., $D'$) be the point such
that $\|D\|= \|C\|$ (resp., $\|D'\| = \|C\|$)
and $O,A,D$ (resp., $O,A',D'$) are collinear.
Then $Sector(O,D,D')$, bounded by the
arc $(D,D')$ centered at $O$, contains either $[A,C]$ or $[A',C']$.
%
%
If it contains $[A,C]$ (see~\refFig{nicely-swept-set}), then we choose
$C''$ such that $[A',C'']\ib Sector(O,D,D')$ and $[A,C]$ is parallel
to $[A',C'']$, and thus \refeq{decomp1} holds.  By symmetry, if the
sector contains $[A',C']$ (see~\refFig{nicely-swept-set-flip}), we can
choose $C''$ so that \refeq{decomp2} holds.
\epf

\noindent {\bf Lemma 5.} {\em
	Let $X=TruncStrip(A,C; A',C'')$.
	There is a basic representation of $X\oplus D(s)$
	of the form $\set{D_1,D_2,D_3,D_4,X'}$ where $D_i$'s
	are discs and $X'$ is the intersection of a convex hexagon
	with an annulus.
} 

\bpf
See \refFig{nicely-swept-set} for a figure of $X$.
Let $D_1=Disc_A, D_2=Disc_C, D_3=Disc_{A'}, D_4=Disc_{C''}$
where $Disc_P$ denote the disc with center $P$ of radius $s$.
These discs are outlined in green in \refFig{Minkowski-sum}.
The boundary of each $D_i$ ($i=1\dd 4$) intersects
the boundary of $X\oplus D(s)$ in a circular arc
$(a_i,b_i)$ where $a_i$ is closer to $O$ than $b_i$.
Let $H_i$ be the half space containing $X$ and bounded by the line
through $[a_i,b_i]$. 
We need to check that these half spaces do indeed contain $X$.
Also, let $H_5$ (resp., $H_6$)
be the half space containing $X$
and bounded by the line through $b_1$ and $a_2$
(resp., $b_3$ and $a_4$). 
Note that $[b_1,a_2]$ and $[b_3, a_4]$ are parallel.
Then we see that $H=\bigcap_{i=1}^6 H_i$ is a convex
hexagon containing $X$, and
the intersection $H\cap Ann(\|A\|-s, \|C\|+s)$
is outlined in red in \refFig{Minkowski-sum}.
Observe that this intersection
covers all of $\Big(X\oplus Disc(s)\Big) \setminus \bigcup_{i=1}^4 D_i$.

This construction is valid as long as $\|A\|\ge s$, i.e.,
the annulus $Ann(\|A\|-s, \|C\|+s)$ is a true annulus.
When $\|A\|<s$, the boundary of $X\oplus Disc(s)$ no longer
has an inner arc of radius $\|A\|-s$, but degenerates into
a concave vertex where the two circles of radius $s$
centered at $A$ and $A'$ (resp.) meet.
\epf


\noindent {\bf Theorem 6.} {\em
	Let $T[\alpha,\beta]$ be a 
	nicely swept set where $[\alpha,\beta]$ has width $\le \pi/2$.
	Then $T[\alpha,\beta]$ can be decomposed into 
	a triangle, a sector and a truncated strip.
	The $s$-expansion of $T[\alpha,\beta]$ has
	a basic representation which is the union of the
	$s$-expansions of the triangle, sector and truncated strip.
} 

\bpf
We know that $T[\alpha,\beta]$ can be decomposed into a triangle
and a swept segment.  The swept segment, since $[\alpha,\beta]$ has
width $\le \pi/2$, can be further decomposed into a sector and a truncated
strip.   The expansions of the triangle and sector are
clear; the expansion of the truncated strip was the subject of the previous
lemma.
\epf


\remove{
%
\noindent {\bf Theorem 5.}
{\em
	$T=[A,B,C]$ is nice if and only if
	for all $\alpha,\beta$ with $[\alpha,\beta]$ less than a half-circle,
        $T[\alpha,\beta]$ is a nicely swept set (NSS).
%
}
%
\bpf
W.l.o.g., assume that $A,B,C$ are in a CCW order (the other case is
symmetric).  Suppose $T$ is nice.  We claim that if $\beta - \alpha >
0$ is small enough, then $T[\alpha,\beta]$ is equal to $(S\cap
Disc(\|C\|)) \setminus Disc(\|A\|)$.
To see this, 
look at~\refFig{nice-tri-general} and the conditions of niceness
in~(\ref{eq:nice}): the first two conditions makes sure that $B \in
H_A$ and $C \in H_A$, i.e., the edge $[A, B]$ (resp.\ $[A,C]$) does
not cut through $Circle A$ (denoting the circle centered at $O$ and
passing through $A$) and thus $T[\alpha,\beta]$ is outside
$Disc(\|A\|)$. Similarly, the third condition makes sure that $C \in
H_B$, i.e., the edge $[B, C]$ does not cut through $Circle
B$. Therefore $[B, C]$ is always above $Circle B$ during sweeping, and
the shape of $T[\alpha,\beta]$ is $(S\cap Disc(\|C\|)) \setminus
Disc(\|A\|)$.
%
%

	Conversely, suppose $T$ is not nice.  Say the first condition
        $\bang{A,B-A}\ge 0$ is violated.  This means that the edge
        $[A,B]$ intersects $Circle A$ at another point $D\in [A,B]$
        (see~\refFig{Not-nicely-swept2}(a)). Consider rotating $[A, B,
          C]$ to $[A', B', C']$ such that $A'$ is between $A$ and
        $D$. Clearly the corresponding swept set $T[\alpha,\beta]$ has
        some portion inside $Disc(\|A\|)$ and thus it cannot be
        $(S\cap Disc(\|C\|)) \setminus Disc(\|A\|)$.  Violating the
        second condition $\bang{A,C-A}\ge 0$ is similar.  Now consider
        violating the third condition $\bang{B,C-B}\ge 0$.  This means
        that the edge $[B,C]$ intersects $Circle B$ at another point
        $D\in [B,C]$ (see~\refFig{Not-nicely-swept2}(b)). Consider
        sweeping from $[A, B, C]$ to $[A', B', C']$ where $B'$ is
        between $B$ and $D$. Then $[B,C]$ and $[B',C']$ intersect at a
        point $E$ below $Circle B$ and we see that the boundary
        portion $[B', E, D]$ of the swept set $T[\alpha,\beta]$ is
        {\bf non-convex}.  Note that a nicely swept set (NSS), before
        subtracting $Disc(\|A\|)$, i.e., $S\cap Disc(\|C\|)$, is {\bf
          convex}, but $T[\alpha,\beta]$ is non-convex even before
        subtracting $Disc(\|A\|)$ (due to $[B', E, D]$), so
        $T[\alpha,\beta]$ is not an NSS.
\epf

	    \begin{figure}[htb]
	    	  \begin{center}
		   \scalebox{0.28}{
	    	     \input{./figs/Not-nicely-swept2.pdf_t}}
	    	   \caption{Proof of Theorem~\protect{\ref{thm-NSS}}:
  The area swept from triangle $[A, B, C]$ to $[A', B', C']$ is not a
  nicely swept set (NSS).}
	    	   \label{fig:Not-nicely-swept2}
	    	  \end{center}
	    \end{figure} 	

} 

\noindent {\bf Lemma 7.}
%
{\em
	If $T$ is an arbitrary triangle and $O$ is exterior to $T$,
	then we can partition $T$ into at most $4$ nice triangles.
}
%

%
\bpf 
%
        Let $T = [A,B,C]$. In the worst case, all three niceness
        conditions for $T$ (i.e., $B \in H_A, C\in H_A$, and $C \in
        H_B$, where $0\le \|A\|\le \|B\|\le\|C\|$; recall the
        geometric view of niceness described right after
        Eq.~\refeq{nice}) are violated. W.l.o.g., suppose that among
        the three edges of $T$, $[A,B]$ is the closest to $O$. Let $D$
        be the point on $[A,B]$ such that $[O,D] \perp [A,B]$, and
        similarly for $E \in [A,C]$ and $F \in [B,C]$;
        see~\refFig{4-nice-triangles}. Then we add segments $[C,D],
        [D,E], [D,F]$ to decompose $T$ into 4 triangles $[D,E,A],
        [D,E,C],[D,F,B]$ and $[D,F,C]$. Note that the line $L_D$
        tangent to the circle of radius $\|D\|$ (centered at $O$) and
        passing through the point $D$ coincides with $[A,B]$;
        similarly, the line $L_E$ coincides with $[A,C]$ and $L_F$
        coincides with $[B,C]$. As before, $H_D$ is the half space
        bounded by $L_D$ and not containing $O$; similarly for $H_E$
        and $H_F$.  For the triangle $[D,E,A]$, note that $0\le
        \|D\|\le \|E\|\le\|A\|$ since $[A,B]$ is closer to $O$ than
        $[A,C]$ (so $\|D\| \le \|E\|$), $[O,D] \perp [D,A]$ (so $\|D\|
        \le \|A\|$) and $[O,E] \perp [E,A]$ (so $\|E\| \le \|A\|$).
        Thus the three niceness conditions for the triangle $[D,E,A]$
        are: $E \in H_D, A \in H_D$, and $A \in H_E$. Again, these
        three conditions are satisfied due to the facts that $[A,B]$
        is closer to $O$ than $[A,C]$, $[O,D] \perp [D,A]$ and $[O,E]
        \perp [E,A]$, i.e., these conditions are automatically
        satisfied due to the construction of $D$ and $E$.  Similarly,
        the three niceness conditions for the triangle $[D,E,C]$ are:
        $E \in H_D, C \in H_D$, and $C \in H_E$, which are again
        satisfied due to the construction of $D$ and $E$.
        Symmetrically, the triangles $[D,F,B]$ and $[D,F,C]$ are both
        nice due to the construction of $D$ and $F$.  Therefore $T$
        can be decomposed into at most 4 nice triangles.
%

%
%
\ignore{
        We refer to~\refFig{4-nice-triangles}. Note that segments
        $[O,D], [O,E]$ and $[O,F]$ are perpendicular to segments
        $[A,B], [A,C]$ and $[B,C]$, respectively. It can be easily
        verified that each of the 4 triangles $[D,A,E], [D,E,C],
        [D,C,F]$ and $[D,F,B]$ is a nice triangle.
} 
    \epf

	    	\begin{figure}[htb]
	    	  \begin{center}
		   \scalebox{0.5}{
	    	     \input{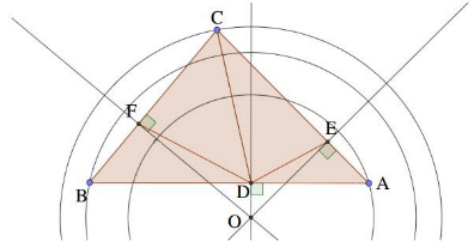}}
	    	   \caption{Proof of
          Lemma~\protect{\ref{lem:4-nice-triangles}}: A triangle $T =
          [A,B,C]$ with the origin $O$ in the exterior can be
          decomposed into at most 4 nice triangles.}
	    	   \label{fig:4-nice-triangles}
	    	  \end{center}
	    	\end{figure} 	  

\noindent {\bf Theorem 8.}
%
{\em
	Let $P$ be an $n$-gon.
	\\(i)
	Given any triangulation of $P$
	into $n-2$ triangles, we can refine the triangulation 
	into a triangulation with $\le 4n-6$ nice triangles.
	\\(ii)
	This bound is tight in this sense: for every $n\ge 3$,
	there is a triangulation of $P$ whose refinement has size $4n-6$.
}

\bpf
	(i) In the given triangulation of $P$, we might
	have a triangle containing $O$.  This triangle can be
	triangulated into at most $6$ nice triangles (Lemma~\ref{lem:star-decompose}).
	By Lemma~\ref{lem:4-nice-triangles},
        the remaining $n-3$ triangles can be 
	refined into $4(n-3)$ nice triangles. 
        The final count is $6+4(n-3)=4n-6$. 
  
	(ii) We construct an $n$-gon whose vertices are all
	on the unit circle.  Note that all such vertices
	are of the form $e^{\ii\theta}$.
	For the first triangle $T_0$, pick the vertices 
	$1, e^{\ii 2\pi/3}, e^{\ii 4\pi/3}$.
%
%
        Call these vertices $u_0, u_1, u_2$. Choose the origin $O$ inside $T_0$
        so that for each triangle $[O, u_i, u_j], i \neq j \in \set{0, 1, 2}$
        we have both $\angle Ou_iu_j < 90\degree$ and $\angle Ou_ju_i < 90\degree$.
        Therefore each triangle $[O, u_i, u_j]$ must be split into $2$ nice triangles
        and overall $T_0$ must be split into $6$ nice triangles.
%
%
%
	If $n=3$, our result is verified.  If $n>3$,
	we must add $n-3$ additional vertices.
	Define the vertex
	$v_k \as e^{\ii \frac{2k\pi}{3(n-2)}}$, $k=0,1\dd n-2$
%
%
	(note that $v_0=1$ and $v_{n-2}=e^{\ii 2\pi/3}$ have
	previously been chosen).  Thus we have added $v_1\dd v_{n-3}$
	new vertices.
	Any triangle $[v_k, v_\ell, v_m]$ must be split into $4$ nice triangles.
	This proves our claim.
	\epf

%

\end{document}